\documentclass[twocolumn]{bytedance_seed}

\usepackage[toc,page,header]{appendix}

\usepackage{minitoc}
\usepackage{xspace}
\usepackage{tikz}
\usepackage{amsmath}
\usepackage{multirow}
\usepackage{filecontents}

\usepackage{subcaption}
\usepackage{makecell}
\usepackage{amssymb}
\usepackage{soul}
\usepackage[algo2e,linesnumbered,algoruled,noend,noline]{algorithm2e}
\usepackage{booktabs}

\newcommand{\allnotes}[1]{}
\newcommand{\ignore}[1]{}

    \newcommand{\newtext}[1]{{#1}}

\newcommand{\eg}{\textit{e.g.,}\xspace}
\newcommand{\ie}{\textit{i.e.,}\xspace}

\newcommand{\company}{ByteDance\xspace}
\newcommand{\Segment}{Operation\xspace}

\newcommand{\segment}{operation\xspace}
\newcommand{\segments}{operations\xspace}
\newcommand{\ndtimer}{NDTimeline\xspace}
\newcommand{\ndtimerNoXSpace}{NDTimeline}

\newcommand{\gcOpt}{planned GC\xspace}

\newcommand{\sysname}{SMon\xspace}
\newcommand{\sysnameNoXSpace}{SMon}

\newcommand{\OpTimeNoXSpace}{OpDuration}
\newcommand{\cpname}{CP\xspace}
\newcommand{\falcon}{FALCON\xspace}

\newcommand{\manyRunsGPUHourRatio}{7.3\%\xspace}
\newcommand{\manyRunsJobRatio}{13.9\%\xspace}

\newcommand{\simGPUHourRatio}{34.1\%\xspace}
\newcommand{\simJobRatio}{50.0\%\xspace}

\newcommand{\largeSimErrorGPUHourRatio}{7.7\%\xspace}
\newcommand{\largeSimErrorJobRatio}{11.2\%\xspace}

\newcommand{\totCovGPUHourRatio}{56.4\%\xspace}
\newcommand{\totCovJobRatio}{38.2\%\xspace}
\newcommand{\totCovJobCnt}{3079\xspace}

\newcommand{\gcExpSpeedup}{12.6\%\xspace}

\newcommand{\stageImbaImprove}{9.9\%\xspace}
\newcommand{\seqlenimbaExpSpeedup}{23.9\%\xspace}

\newcommand{\stageImbaFwdRatioBeforeOpt}{2.07\xspace}
\newcommand{\stageImbaBwdRatioBeforeOpt}{1.41\xspace}
\newcommand{\stageImbaFwdRatioAfterOpt}{1.55X\xspace}

\newcommand{\blockingWaste}{10.4\%\xspace}

\newcommand{\averageBlockingSldSJ}{1.28\xspace}

\newcommand{\fixedHostBlockingSld}{3.04\xspace}
\newcommand{\fixedHostBlockingRatio}{1.7\%\xspace}

\newcommand{\fixedGroupBlockingRatio}{39.3\%\xspace}

\newcommand{\PNineZeroWaste}{21.3\%\xspace}

\newcommand{\PNineNineWaste}{45.0\%\xspace}

\newcommand{\straggleJobRatio}{42.5\%\xspace}

\newcommand{\noPPJobRatio}{21.1\%\xspace}
\newcommand{\seqlenimbaJobRatio}{21.4\%\xspace}
\newcommand{\seqlenimbaSlowdown}{1.34\xspace}

\newcommand{\mildAct}{1.16\xspace}
\newcommand{\mildEst}{1.21\xspace}

\newcommand{\mediumAct}{1.40\xspace}
\newcommand{\mediumEst}{1.42\xspace}

\newcommand{\severeAct}{2.03\xspace}
\newcommand{\severeEst}{1.98\xspace}

\title{Understanding Stragglers in Large Model Training Using What-if Analysis}

\author[1]{Jinkun~Lin}
\author[2]{Ziheng~Jiang}
\author[2]{Zuquan~Song}
\author[2]{Sida~Zhao}
\author[2]{Menghan~Yu}
\author[1]{Zhanghan~Wang}
\author[2]{Chenyuan~Wang}
\author[4]{Zuocheng~Shi}
\author[3]{Xiang~Shi}
\author[2]{Wei~Jia}
\author[2]{Zherui~Liu}
\author[2]{Shuguang~Wang}
\author[2,\dagger]{Haibin~Lin}
\author[2]{Xin~Liu}
\author[1]{Aurojit~Panda}
\author[1,\dagger]{Jinyang~Li}

\affiliation[1]{New York University}\affiliation[2]{ByteDance Seed}\affiliation[3]{ByteDance}\affiliation[4]{Zhejiang University}

\contribution[\dagger]{Corresponding authors}

\abstract{
Large language model (LLM) training is one of the most demanding distributed computations today, often requiring thousands of GPUs with frequent synchronization across machines. Such a workload pattern makes it susceptible to stragglers, where the training can be stalled by few slow workers. At \company we find stragglers are not trivially always caused by hardware failures, but can arise from multiple complex factors. This work aims to present a comprehensive study on the straggler issues in LLM training, using a five-month trace collected from our \company LLM training cluster. The core methodology is what-if analysis that simulates the scenario without any stragglers and contrasts with the actual case. We use this method to study the following questions: (1) how often do stragglers affect training jobs, and what effect do they have on job performance; (2) do stragglers exhibit temporal or spatial patterns; and (3) what are the potential root causes for stragglers?
}

\date{\today}
\correspondence{Haibin Lin at \email{haibin.lin@bytedance.com}, Jinyang Li at \email{jinyang@cs.nyu.edu}}

\checkdata[Project Page]{\url{https://github.com/ByteDance-Seed/StragglerAnalysis}}

\begin{document}
\maketitle

\section{Introduction}\label{sec:introduction}
Large-language models (LLMs) have been widely adopted and are being used for a variety of tasks. In response to their adoption, many companies are focused on training ever-larger models because empirically an increase in model size has been shown to improve task accuracy. For example, in little more than a year, Meta has moved from the 65 billion parameter Llama-1 model to the 405 billion parameter Llama-3 model. Training these large models in a reasonable time requires using resources, including GPUs and memory, distributed across several thousand servers in a cluster. However, unlike previous distributed parallel data processing workloads such as MapReduce, distributed LLM training requires frequent synchronization and tighter coordination. Consequently, slow workers or \emph{stragglers} can significantly hurt training time and resource efficiency. This paper seeks to answer the question \textit{do stragglers actually pose a serious performance issue in real-world large-scale LLM training deployment?} 

The impact of stragglers in LLM training is dictated by the parallelization strategy used to distribute LLM training across a cluster. A typical LLM training job uses a hybrid parallelism strategy that combines multiple parallism strategies, including
pipeline parallelism (PP)~\cite{narayanan2021efficient,huang2019gpipe,fan2021dapple} and data parallelism (DP)~\cite{rajbhandari2020zero} to partition state and computation across servers. All of these strategies require frequent coordination to synchronize results across GPUs and servers, and are thus vulnerable to stragglers. A microbatch that is delayed due to a straggler in a PP stage can slow down the entire training batch in DP.
Further, few approaches exist to mitigate the effect of stragglers in LLM training: traditional straggler-mitigation approaches~\cite{ananthanarayanan2010reining, ananthanarayanan2013effective,tensorflow} that rely on backup workers assume infrequent synchronization and would significantly hurt training performance and resource requirements for LLM training jobs. On the other hand, approaches that use asynchronous SGD~\cite{dean2012large} or drop updates from slow workers~\cite{giladi2023dropcompute} change training behavior, and are not widely adopted due to concerns about their effect on model accuracy.

Given that LLM training jobs are vulnerable to stragglers, we set out to understand their effect on real-world LLM training clusters. To do so we analyzed traces collected from the LLM training cluster at \company over a five-month period from January to May 2024 (\S\ref{subsec:trace-collection}). Our analysis seeks to address the following questions:
\begin{itemize}
    \item How often do stragglers affect training jobs? What effect do they have on training job performance?
    \item Do stragglers exhibit temporal or spatial patterns? For example, do they appear in most or only a few steps of a training job? Are they confined to a few workers or more widespread across workers?
    \item What are the potential root causes for stragglers? What is the relative severity of different root causes?
\end{itemize}

We answer these questions using ``what-if'' analysis (\S\ref{subsec:what-if}): we estimate how a training job would perform if all (or a subset) of stragglers had not occurred. 
To do so, we identify the \emph{dependency model} within a training job's traced \segments, and then perform a simulation based on each \segment's non-straggling execution time to quantify job-completion time in the absence of stragglers.

Our analysis revealed that stragglers are widespread and have notable performance impacts on LLM training; \straggleJobRatio of the jobs are at least 10\% slower due to stragglers. At the tail, stragglers can result in jobs wasting 45\% of their allocated resources (\S\ref{subsec:general-impact}). Most steps incur similar slowdowns within a straggling job, suggesting that they are often not caused by transient environmental issues but are rather caused by persistent problems (\S\ref{subsec:steps}). Our analysis further shows that stragglers are more commonly caused by a slowdown in computation \segments rather than communication \segments (\S\ref{sec:optype-impact}). We did not notice a positive correlation between job size and its straggler-related slowdown, suggesting factors other than job size play a dominant role in straggling (\S\ref{subsec:jobsize}).

Coupling simulation-based analysis with manual inspection, we investigate several root causes for straggling. We found that hardware or software problems in a server were not a significant cause of the stragglers we observed (\S\ref{sec:machine-issue-root-cause}). On the other hand, imbalance in how work was partitioned across pipeline stages (\S\ref{sec:fixed-group-root-cause}), an imbalance in sequence lengths included in each microbatch (\S\ref{sec:seqlenimba-root-cause}), and pauses induced by the garbage collector (\S\ref{sec:gc-root-cause}) were responsible for many of the stragglers we observed. We summarize the key observations and their implications in \S\ref{sec:new-obs}.

Our analysis pipeline also suggests an approach to make it easier to detect and debug stragglers when a training job is running. As a result, we incorporated portions of our analysis pipeline into a monitoring system called \emph{\sysnameNoXSpace} (\S\ref{sec:deployment}) that is now deployed in the training cluster at \company. \sysname partially automates the manual analysis we report on in this paper, and is used by the on-call team at \company to detect and address stragglers in important jobs.

\section{Background}
\label{sec:background}
We start by providing background on distributed LLM training and sources of stragglers.

\subsection{Hybrid-Parallel Training and Stragglers}

Several different strategies have been developed to parallelize LLM training to overcome memory limitations and reduce training time. Each parallelization strategy is susceptible to stragglers. Below, we introduce the strategies used in our cluster and the sources of stragglers. We describe each strategy in terms of how computation and memory are partitioned across \emph{workers}, a term we use to refer to a single GPU and a process (running on the CPU) that controls it.

\paragraph{Data parallelism (DP), ZeRO~\cite{rajbhandari2020zero} and fully sharded data parallelism (FSDP~\cite{zhao2023pytorchfsdp})} DP partitions training data across multiple workers, each of which has a replica of the entire model. When DP is used, in each training step, a worker is assigned a training batch. The worker runs forward and backward computation for its assigned batch, and then all workers perform a gradient all-reduce step before moving to the next training step. The gradient all-reduce step requires synchronization between workers, and a slow worker can cause all workers to stall, leading to stragglers. ZeRO~\cite{rajbhandari2020zero} and FSDP~\cite{zhao2023pytorchfsdp} extend DP to reduce the per-GPU memory demand by partitioning the optimizer state, parameters, and/or gradients across workers. Instead of using a gradient all-reduce step, both ZeRO and FSDP require a reduce-scatter step that computes gradients, a device-local parameter update step, and an all-gather step to collect parameters for each training step. The reduce-scatter and all-gather steps require synchronization across workers, and are thus similarly susceptible to stragglers.

\paragraph{Pipeline parallelism (PP).}  
PP partitions the model across multiple workers, each of which holds a disjoint set of consecutive model layers, referred to as a \emph{pipeline stage}. PP reduces the per-GPU memory demand for model weights and activations.  During training, a batch of data is split into several microbatches, and training is pipelined through PP stages. Several microbatch scheduling approaches have been proposed, including GPipe~\cite{huang2019gpipe}, 1F1B~\cite{fan2021dapple}, and virtual pipeline parallelism (VPP)~\cite{narayanan2021efficient}. All of these scheduling approaches assume that computation is evenly partitioned across pipeline stages, and aim to minimize \emph{pipeline bubbles}, \ie times when a pipeline stage is idle waiting for data from the previous stage. If PP stages are not evenly partitioned, the slowest stage stalls other stages and becomes a performance bottleneck. Thus, PP is susceptible to straggling due to uneven compute partitioning between PP stages. For ease of exposition, we do not explicitly discuss VPP, but our analysis does account for it.

\paragraph{Tensor parallelism (TP) and context parallelism (CP).}
In addition to PP, one can use Tensor (TP)~\cite{narayanan2021efficient,korthikanti2023reducing} and context  (CP)~\cite{contextparallel,dubey2024llama,liu2023ringattentionblockwisetransformers} parallelism to further reduce the per-GPU memory demand. TP partitions each layer's weight between workers, and CP partitions each sequence's tokens across workers. Both require a synchronization step after each transformer layer so that partial results from all workers in the TP or CP group can be aggregated.
As a slow device can stall progress during synchronization, both TP and CP are susceptible to stragglers. As we discuss in \S\ref{s:limitations}, we do not analyze systematic stragglers in the TP and CP groups.

\begin{figure}
    \centering
    \includegraphics[width=1\linewidth]{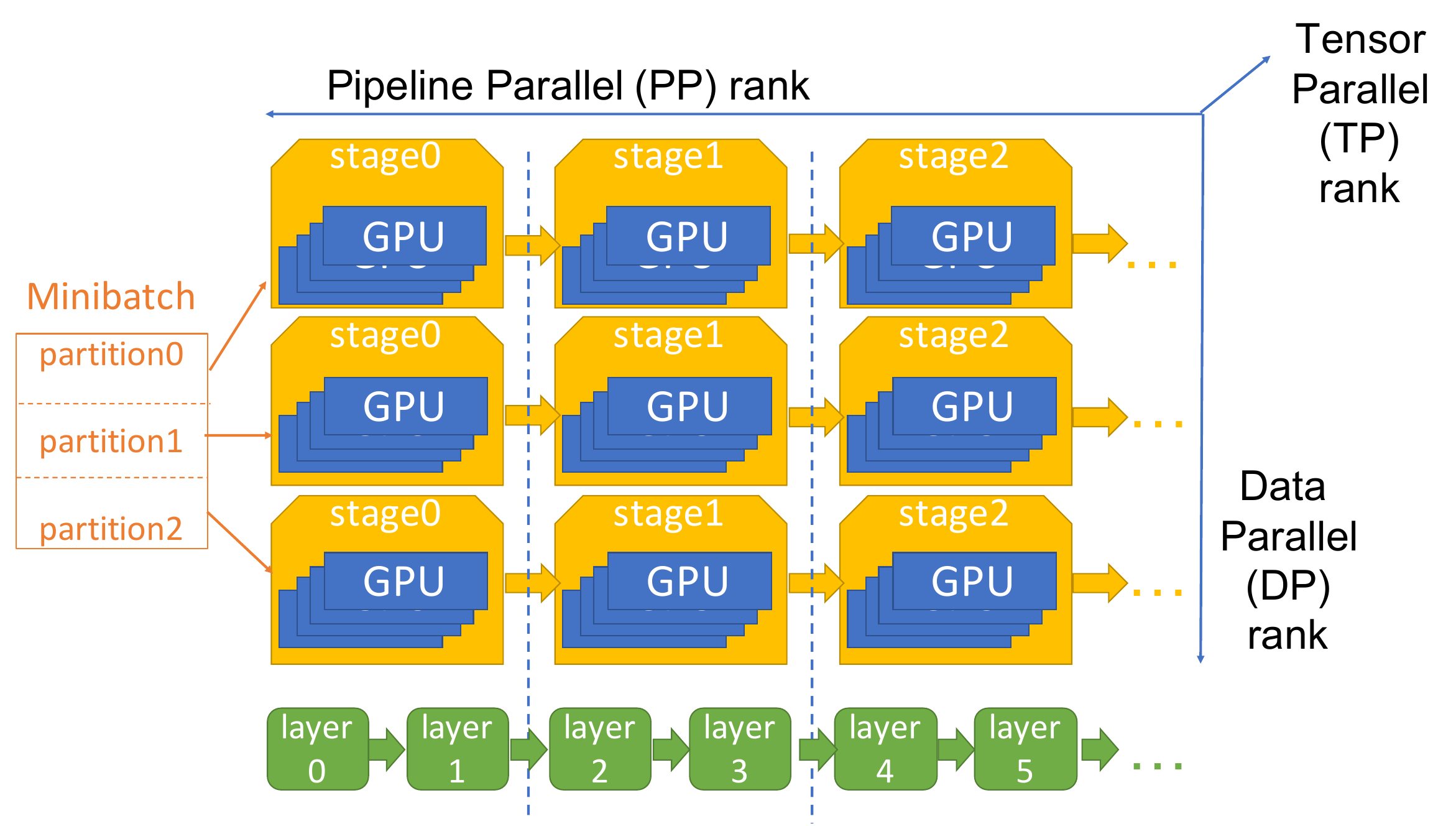}
    \caption{Rank topology of a DP-PP-TP parallelism.}
    \label{fig:3d-parallel}
\end{figure}

\paragraph{Hybrid parallelism.} In practice, LLM training uses a hybrid strategy that combines all of the parallelism strategies discussed above and provides better performance than any individual strategy.\footnote{Other parallelisms such as expert parallelism (EP)~\cite{lepikhin2020gshard} also exist but are not used in the jobs that we analyze.}
When hybrid parallelism is used, workers can be organized into a hypercube, where each dimension corresponds to one parallelism strategy. A worker's coordinate provides its \emph{rank} (\eg pipeline stage) in each parallelism dimension. Workers are also assigned a unique \emph{global-rank} for identification.
Figure~\ref{fig:3d-parallel} shows an example worker topology with DP-PP-TP parallel training. 

Training with hybrid parallelism operates in a layered fashion.
For example, when DP-PP-TP parallelism is used, workers with the same DP rank are grouped into a single PP group, which is responsible for computing gradients for a training batch. Inside each PP group, workers with the same PP rank are grouped into a TP group, each of which is responsible for a single PP stage. Finally, workers within the same TP group use TP parallelism to compute each layer in the PP stage.

\paragraph{The case of straggling for hybrid-parallelism.} Hybrid-parallelism is affected by stragglers appearing in any of the parallelism strategies that it combines. For example, in the DP-PP-TP parallelism strategy, a slow TP worker will slow down the PP rank to which it belongs, leading to a PP bubble. This PP bubble, in turn, will slow down the entire DP rank to which the worker belongs, delaying gradient synchronization and stalling other DP ranks.

\subsection{Goals and Challenges}
\label{subsec:goal}

\newtext{\paragraph{Straggler definition.} An LLM training job that uses hybrid parallelism is straggler-free if all workers take the same amount of time to complete their assigned work. 
This minimizes the time required for synchronization, resulting in an ideal straggler-free scenario that achieves the best possible performance.
We also note that, by this fairly broad definition of straggling, a worker lagging behind others for any reason is considered a straggler.   These reasons can include not only hardware issues that deterministically affect one single worker, 
but also unpredictable stalls that affect all workers uniformly (\eg garbage collection that prevents a task from making progress~\cite{imbue,jiang2024megascale}), 
as well as workload imbalance due to data skew~\cite{ousterhout2015making,zhang2024disttrainaddressingmodeldata} or poor workload partitioning across weights (\eg PP stage)~\cite{dubey2024llama}.
}

\newtext{\paragraph{Our goal.} In this paper, we want to investigate how straggling impacts real-world LLM training jobs and to unearth some of the underlying root causes.  To do so, we seek to quantify \emph{how much an actual LLM training job's speed differs from that of some ideal straggler-free case}.  In keeping with our earlier definition, a straggling worker is one that completes its work more slowly than others. Thus, in the presence of stragglers, a job is delayed by its need to synchronize across groups of workers, resulting in a slower speed compared to the ideal straggler-free scenario. 

Our study is based on analyzing traces from LLM training jobs collected over a five-month period. Although these traces capture each job’s actual execution time, estimating the corresponding straggler-free completion time remains challenging. This is because it is tricky to assess straggling operations' impact on the overall job duration in the presence of overlapping execution. Traditional critical path analysis~\cite{hta} falls short in this context, as highly parallel and homogeneous workloads like LLM training can exhibit many similarly critical paths. Focusing on a single path can lead to misleading conclusions, as shown in Coz~\cite{curtsinger2015coz}. To address this, we instead use trace-based simulation to "execute" each job on an alternative timeline where straggling operations are brought in line with their peers.}

\section{Methodology}
\label{sec:method}

In this section, we describe the traces used for our study and our methodology for analyzing them.  

\subsection{The LLM Training Job Traces} 
\label{subsec:trace-collection}

\paragraph{The cluster setup.} The cluster on which the traces have been collected is dedicated for training and shared internally by multiple teams.  The machines used in the cluster have a similar hardware configuration as NVIDIA's DGX servers: each server is equipped with eight GPUs interconnected with NVLink or PCIe links, four or eight NICs with hundreds of Gbps bandwidth, a separate NIC used for storage and management, a few hundred CPU cores and a few TBs of memory. Servers are interconnected through high-performance switches configured in a three-layer CLOS topology.  The network is overprovisioned and carefully tuned~\cite{jiang2024megascale} to ensure that there are no slowdowns due to network congestion.

Multiple jobs can be scheduled to run on the cluster at the same time, with each job having exclusive use of its allocated GPUs for the duration of the job. The scheduler ensures that each job uses homogeneous hardware by allocating it the same type of GPUs.  Furthermore, the scheduler performs the GPU allocation in a best-effort manner to ensure a job's GPUs are close together in the network topology.  Since large jobs request GPUs in multiples of eight, different large jobs do not share the same server machine.  Combined with the lack of network congestion, this means we do not see stragglers due to resource contention despite jobs running on the same cluster.

\begin{table*}[t]
    \centering
    \resizebox{\textwidth}{!}{%
    \begin{tabular}{lll}
    \toprule
      & Type of \Segment &  Description \\ \cmidrule(lr){1-3}
    \multirow{2}{*}{Compute Op.} & forward-compute  & Forward computation done for a microbatch for a PP stage.\\
    & backward-compute & Backward propagation done for a microbatch for a PP stage.\\\cmidrule(lr){1-3}
PP-specific & forward-send, forward-recv & P2P communication between PP stages for a microbatch on the forward pass. \\
communication Op.& backward-send, backward-recv & P2P communication between PP stages for a microbatch on the backward pass. \\ \cmidrule(lr){1-3}
\multirow[c]{2}{*}{\makecell[l]{\\DP-specific \\communication Op.}} & params-sync& \makecell[l]{Collective communication (all-gather) among all DP ranks for a given PP stage\\ to retrieve the stage's weights prior to the first microbatch's forward-compute.} \\ \cmidrule(lr){2-2} \cmidrule(lr){3-3}
& grads-sync & \makecell[l]{Collective communication (reduce-scatter) among all DB ranks for a given PP stage\\ to aggregate the stage's gradients after the last microbatch's backward-compute.} \\ 
   \bottomrule
    \end{tabular}}
    \caption{Types of \segments traced by \ndtimer and their description.
    }
    \label{tab:op-type-def}
\end{table*}
\paragraph{Trace collection.} The traces used in our study are collected from LLM pretraining jobs submitted during the time period from 2024/01/01 to 2024/05/31. As our study focuses on large jobs, we only use traces for jobs that use at least 128 GPUs. \newtext{We also discard traces that are invalid or not suitable for analysis as detailed in \S\ref{s:limitations}.} This produces \totCovJobCnt
jobs for our analysis.  These jobs contain both dense and mixture-of-experts (MoE) models that are configured to train using either short or long context sequences. 
\newtext{Of the $3079$ jobs, $31.7\%$ use $\geq256$ GPUs, $18.3\%$ use $\geq512$ GPUs, and $3.6\%$ use $\geq 5000$ GPUs. In total, the analyzed jobs cover around half of the total GPU-hours among all LLM training jobs.}

All jobs in our traces are done using customized versions~\cite{jiang2024megascale} of the open-source distributed LLM training system, Megatron-LM~\cite{megatron}. The jobs are configured to use different combinations of the parallelization strategies supported by Megatron-LM including DP, PP, TP and CP. \footnote{CP in our traces uses an in-house implementation with a slightly different algorithm than the CP in open-source Megatron-LM.}

The Megatron-LM based training system has been instrumented with our in-house profiling tool \emph{\ndtimerNoXSpace}~\cite{ndtimeline}.  By default, it samples $10\%$ of the training steps (aka iterations) of a job for profiling.  For each profiled step, the tool records the start and end time of a set of operations that are deemed important, including those performing forward and backward computation as well as those performing communication.  Table~\ref{tab:op-type-def} gives all the profiled \emph{\segment types} and their description.
A profiled forward or backward compute operation consists of many GPU kernels in order to reduce the cost of tracking many small computation kernels.   
\ndtimer periodically synchronizes the clocks of all machines for a job, thereby allowing us to align related operations across different machines for the purpose of what-if analysis (\S\ref{subsec:what-if}).

For each operation in the trace, its logged entry contains the operation type, its start and end timestamps as well as a set of metadata
including the training step ID, the microbatch ID, the PP rank (within each PP group) and the DP rank (within the overall DP group).
These metadata enable us to reconstruct the operation dependencies which are necessary in order to simulate an alternative timeline had stragglers not been present (\S\ref{subsec:what-if}).

\subsection{Simulator for What-if Analysis}
\label{subsec:what-if}

The goal of the what-if analysis is to assess the impact of stragglers by answering the following questions:
\begin{enumerate}
    \item \emph{How long would a job take if all stragglers were absent?}
    \item \emph{How long would a job take if all except for a certain group of stragglers were absent?}
\end{enumerate}

To answer these questions, we simulate an alternative timeline without stragglers. The main insight is that comparable operations would have the same duration in the absence of stragglers. Based on this insight, we first try to estimate the duration of each operation in the alternative straggler-free scenario. Then, we simulate an alternative timeline in which operations launch according to their dependencies and complete within their estimated idealized durations. By comparing the simulated job completion time (JCT) to that of the actual trace, we can assess the impact of stragglers.

\paragraph{Estimate idealized operation durations for straggler-free scenarios.}  Conceptually, we can organize the traced operations into a multi-dimensional tensor with four dimensions: training step, microbatch, PP rank and DP rank. Let's refer to this tensor as an {\em \OpTimeNoXSpace} tensor. We have one such tensor for each operation type in Table~\ref{tab:op-type-def}. For compute operations, the elements in the tensor are simply the corresponding operation duration in the trace.  For communication, we calculate a sub-part of the traced duration called \emph{transfer-duration}.  As communication is done as part of a collective (or a P2P pair), the traced duration of an individual operation is affected by two factors: 1) the amount of time to transfer the data to another rank, aka the ``transfer-duration''.  2) the amount of time to wait for the other operations in the collective (or P2P pair) to start, aka the ``blocking-duration''. Of the two factors, ``transfer-duration'' is intrinsic to the collective operation while the ``blocking-duration'' is determined by operation scheduling.  Therefore,
we only store the ``transfer-durations'' in the {\em \OpTimeNoXSpace} for a communication operation type, e.g. {\em params-sync}, {\em forward-send} etc. To estimate an operation's ``transfer-duration'', we take the maximum start time among all of its peer operations in the same collective (or the same P2P pair) and subtract it from its end time.

All operations of the same type handle the same workload, implying that, in the absence of stragglers, all elements of the idealized {\em \OpTimeNoXSpace} tensor would be equal. To assess the impact of a particular group of stragglers—such as those specific to a machine with PP-rank $p$ and DP-rank $d$—we selectively ``fix'' straggling on all other machines by overriding the duration of operations on those machines with their idealized values.
For those elements belonging to the straggling machine (aka \OpTimeNoXSpace[:,:,p,d]), we keep their original durations. 
We adopt a similar approach (in \S\ref{sec:root-cause}) to estimate the effect of only fixing some straggling elements: we use idealized operation times only for the elements that need to be fixed, and leave the others unchanged.

Once we determine the parts of {\em \OpTimeNoXSpace} tensor that should have the same idealized values in a hypothetical straggler-free setup, the question remains what that value should be.  
\newtext{We use different methods for computation and communication operations.
For the computation operation, we use the average value across the group of elements to be equalized.  For communication operations, we use median instead of average. Initially, we used average for both types of operations, but changed our solution after some manual root cause investigation. In particular, we observe that computational straggling arises predominantly due to workload partitioning imbalance. Therefore, taking the average makes sense as doing so amounts to workload re-balancing.  On the other hand, communication operations have identical transfer volumes across different training steps, microbatches, PP ranks and DP ranks, but they straggle due to external issues like switch/NIC flapping. Furthermore, the affected operations tend to be very long, which significantly skews the average value and makes median a better choice. }

\paragraph{Extract operation dependencies.} 
 
Our simulator requires two pieces of input: the idealized \segments durations and the operation dependency model.  We have previously discussed how to estimate idealized operation duration. Next, we will explain the simulator's dependency model, which is derived from the Megatron-LM based training system in use.

\begin{figure}[t]
\centering
\includegraphics[width=\linewidth]{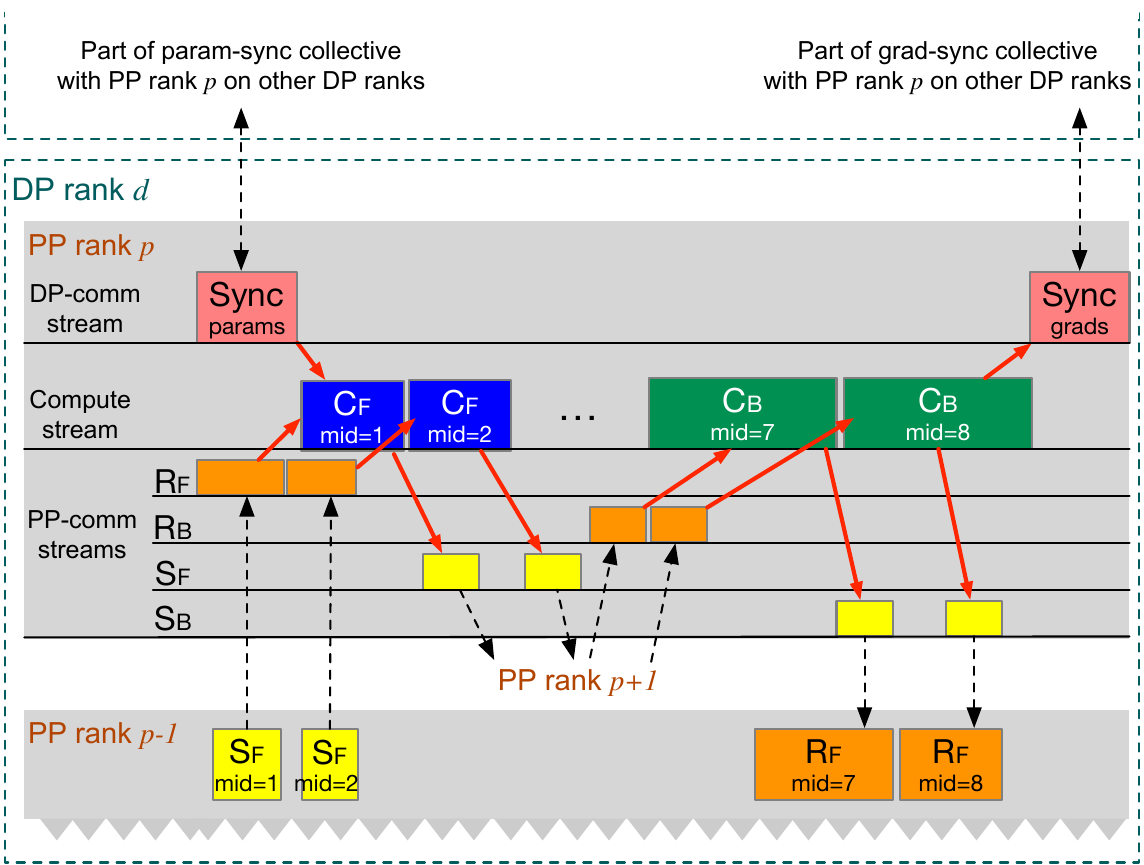}
\caption{The dependency model used in simulation. ``mid'' denotes microbatch ID. Operations within a stream execute sequentially.  Across streams on the same rank, receive operations (\eg $R_{F}$, $Sync_{params}$) precede the corresponding compute operation requiring the data. The opposite dependencies exist for compute and send operations.  Across DP ranks, params-sync ($Sync_{params}$) and grads-sync ($Sync_{grads}$) on the same PP rank form a collective group. Across two adjacent PP ranks, their send and receive operations form a P2P pair. }
\label{fig:megatron-dep}
\end{figure}

In this dependency model, each worker runs several ``streams'' to execute its operations. All operations scheduled to run on a single stream execute sequentially, while operations across streams execute concurrently as long as their dependencies are satisfied.  More concretely, each worker has a stream executing all forward and backward compute operations, a stream executing all DP-specific communication operations and four streams each executing a different type of PP-specific communication operations: forward-recv ($R_F$), backward-recv ($R_B$), forward-send ($S_F$) and backward-send ($S_B$), as shown in Figure~\ref{fig:megatron-dep}.  The dependencies among operations are as follows:
\begin{itemize}
\item \emph{Same stream dependency:} Operations inside a stream are sorted according to their launch time in the trace. We assume an implicit dependency between adjacent operations since operations on the same stream execute sequentially.
\item \emph{DP communication and compute dependency:} The first microbatch's forward-compute operation at each PP stage should happen after the corresponding params-sync collective communication to fetch that stage's parameters, as shown in Figure~\ref{fig:megatron-dep} ($Sync_{params}\rightarrow C_{F,mid=1}$). The parameters are cached locally and used for subsequent microbatches.  The gradients computed for different microbatches are locally accumulated and then aggregated across DP ranks. Thus, the last microbatch's backward-compute should happen before the grads-sync collective communication to aggregate the gradients of that PP stage across DP ranks, as shown in Figure~\ref{fig:megatron-dep} ($C_{B,mid=8} \rightarrow Sync_{grads}$).
\item \emph{PP communication and compute dependency:} Except for the first PP rank, the forward- and backward-compute operation of a microbatch on PP rank $p$ must start after the completion of that microbatch's forward- and backward-receive communication operation on the same GPU, as shown in Figure~\ref{fig:megatron-dep} (e.g. $R_{F,mid=1}\rightarrow C_{F,mid=1}$, $R_{B,mid=7}\rightarrow C_{B,mid=7}$).   Similarly, except for the last PP rank, the forward- and backward-send operation of a microbatch on PP rank $p$ must start after the completion of that microbatch's forward- and backward-compute operation on the same GPU, as shown in Figure~\ref{fig:megatron-dep} (e.g. $C_{F,mid=1}\rightarrow S_{F,mid=1}$, $C_{B,mid=7}\rightarrow S_{B, mid=7}$).
\item \emph{Cross-rank communication dependencies:} The DP  communication operations (aka params-sync, grads-sync) for a given microbatch form a collective group among all DP ranks with the same PP rank.  Similarly, the PP send and receive operations for a given microbatch form a pair between adjacent PP ranks with the same DP rank.  The dependency model for a group of collective (or P2P) operations is that no individual operation can start its data transfer until all operations have been launched.   
\end{itemize}

\paragraph{Simulate an alternative timeline.} 
Using the dependency model and idealized durations, we can simulate an alternative execution timeline using the following rules:
\begin{itemize}
    \item The simulator launches an \segment as soon as all of its dependent \segments finish. In other words, an \segment's start time is calculated as the maximum end time of its dependent \segments.
    \item A computation \segment after launch is immediately marked as finished with end time calculated as the start time plus the corresponding \segment duration in {\emph \OpTimeNoXSpace}.
    \item For each communication \segment, the simulator waits for all its peer \segments in the same collective group (or P2P pair) to launch. An \segment's end time is calculated as the maximum launch time among the group plus the corresponding transfer-duration stored in {\em \OpTimeNoXSpace}. 
\end{itemize}

\subsection{Metrics for Straggler-related Slowdown and GPU Waste }\label{sec:gen:what-if}

Now that we have designed a simulator to estimate a job's completion time in alternative straggler-free timelines, what metrics should we calculate to quantify the impact of stragglers?  Let us use $T_{ideal}$ to denote the JCT without stragglers. Similar to \cite{ousterhout2015making}, to account for errors introduced from simulation, we also simulate the original timeline using unmodified operation durations and denote the resulting JCT as $T$.  The simulation errors are relatively small and reported in \S\ref{sec:validate}. Furthermore, a small number of our traces had large simulation errors, and to ensure analysis fidelity, we discard traces with simulation error $\ge 5\%$ (\S\ref{sec:validate}).
To quantify straggler-related performance slowdown, we calculate the slowdown metric $S$ as a ratio: 
\begin{equation}
\label{eq:blocking-slowdown}
    S\triangleq T / T_{ideal}.
\end{equation}

In addition to the overall straggler-related slowdown, we also want to quantify slowdowns due to straggling in different types of \segments (Table~\ref{tab:op-type-def}).  To do so, we first compute the \emph{\segment-type slowdown} $S_t$ for each \segment-type $t$ as
\begin{equation}
    S_{t}\triangleq T_{ideal}^{-t} / T_{ideal}.
\end{equation}
where $T_{ideal}$ is the ideal JCT computed as described above, and $T_{ideal}^{-t}$ is the JCT when elements in {\em \OpTimeNoXSpace} for \segment-type $t$ are not fixed.

In our cluster, a job has exclusive access to a GPU for its entire duration. As a result, an increase in job completion time (or a slowdown) can be directly translated to the amount of GPU-hours wasted by the job. Concretely, we estimate a job's {\em resource waste} as the percentage of GPU-hours wasted using the equation:
\begin{equation}
\label{eq:total-waste}
    \frac{T-T_{ideal}}{T}=1-\frac{1}{S}.
\end{equation}

Similarly, we can calculate wasted resources due to different \segments types by computing $1-1/S_t$.

\section{Impact of Stragglers}
\label{sec:general-impact}
We perform simulation-based what-if analysis on the job traces and report the impacts of stragglers on our cluster.

\subsection{Stragglers are prevalent and cause non-negligible resource waste.}
\label{subsec:general-impact}

\begin{figure}[t]
    \centering
    \includegraphics[width=.9\linewidth]{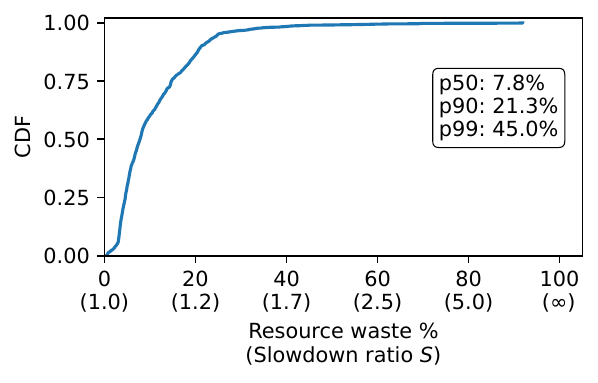}
    \caption{CDF of resource waste among all jobs. A given waste percentage's corresponding slowdown ratio is shown in parentheses.}
    \label{fig:waste-distribution}
\end{figure}
\begin{figure}[t]
    \centering
    \includegraphics[width=.9\linewidth]{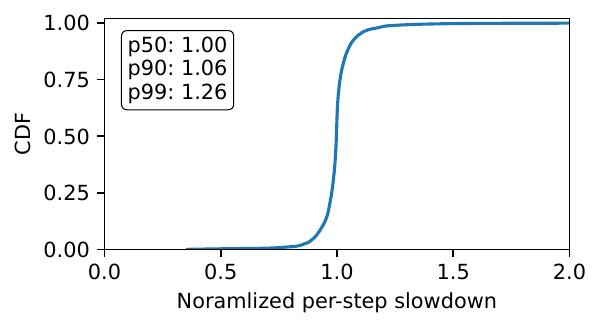}
    \caption{CDF of per-step slowdowns normalized by job slowdown for 15 randomly selected steps from each straggling job.}
    \label{fig:step-slowdown-dist}
\end{figure}

Figure~\ref{fig:waste-distribution} plots the cumulative distribution function (CDF) of the resource waste percentage among all jobs.
We observe that \straggleJobRatio of the training jobs that we trace straggle. Furthermore, because of stragglers, $>10\%$ of the jobs waste at least \PNineZeroWaste of the GPU-hours it is allocated, while $\sim 1\%$ jobs waste at least \PNineNineWaste of its allocated GPUs. Across all our traces we found that \blockingWaste of the allocated GPU hours are wasted due to stragglers.

We also investigated jobs with large slowdowns ($S>3$), and found that all of them were large jobs, 
where fewer than 3\% of the workers were responsible for the problem. Furthermore, in most cases, the slow \segments were performing computation, rather than communication. From this, we hypothesize that server problems (which could be problems with the hardware or misconfiguration) are generally to blame for these problems.

\subsection{Steps have similar slowdowns in a straggling job.}
\label{subsec:steps}

Next, we investigate whether a few really slow steps or most steps contribute to a straggling job's slowdown. We refer to a job as straggling if it has a significant slowdown ratio $S>1.1$.
To do so, we define a step's slowdown as the ratio of the step's execution time to the ideal step execution time (\ie $\frac{T_{ideal}}{n}$ for a job with $n$ training steps).
Figure~\ref{fig:step-slowdown-dist} shows the CDF of per-step slowdown normalized by the job's overall slowdown ratio. We observe that the median step has a normalized slowdown of $1.0$, and even the 90th percentile step has a normalized slowdown of only $1.06$. This indicates that most steps have a slowdown similar to the job's overall slowdown, suggesting that stragglers are not caused by temporary environmental factors but by more persistent problems.
This result also implies that it is sufficient and cost-effective to sample a few training steps to profile stragglers.

\subsection{Which \segment types are to blame for straggling?}
\label{sec:optype-impact}
\begin{figure}[t]
    \centering
    \includegraphics[width=.9\linewidth]{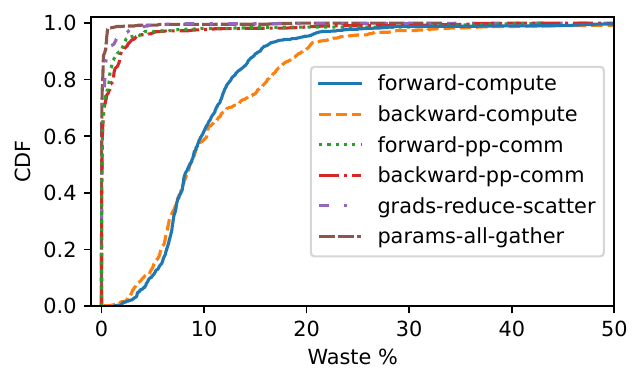}
    \caption{Resource wasted due to the slowdown of a single type of \segment.
    Computation slowdown caused the most resource waste, while communication had a minimal impact.
    }
    \label{fig:slow-seg-ty-contrib}
\end{figure}

Figure~\ref{fig:slow-seg-ty-contrib} shows the fraction of resource waste by each \segment type across all jobs in our traces.
In our traces, communication time measures time spent transferring data, and thus a slowdown in send times produces a corresponding slowdown in receive times. As a result, we group both together in the figure.
Contrary to the observation in \falcon~\cite{wu2024falcon} we find that most slowdowns are caused by computation \segments, as opposed to communication.
This is owing to our ample network bandwidth, the use of dedicated LLM-training clusters, and several in-house network optimizations and tunings~\cite{jiang2024megascale}.

Figure~\ref{fig:slow-seg-ty-contrib} also shows that PP-level communication has a slightly higher impact than the DP-level communication \segments.
This matches our expectation since DP-level communication is overlapped by a lot and can tolerate more slowdowns than PP-level communication kernels, many of which happen in the warmup and cooldown phases, which are on the critical path of a training step.

\subsection{How does job size correlate with straggling?}
\label{subsec:jobsize}

Surprisingly, we do not observe an obvious positive correlation between the slowdown and job size. 
This suggests that job size is not the determining factor of straggling, and other factors like the model type or human factors may play a more important role.
For instance, very large jobs are usually babysat by our on-call team and optimized better than others. Therefore, their slowdowns are not necessarily worse than those for smaller jobs.
As another example, long-context jobs are observed to suffer more from stragglers (\S\ref{sec:seqlenimba-root-cause}), but they typically have smaller job sizes, which when mixed with other models biased the result towards the opposite job size and slowdown correlation.

\section{Root Causes}
\label{sec:root-cause}

In this section, we analyze the traces to diagnose the root causes of stragglers. Our analysis focuses on straggling jobs, aka those with slowdown $S\ge 1.1$. Identifying the root cause of a job slowdown requires manual examination, and thus we cannot investigate all possible root causes. Instead, we focus on common root causes, i.e., ones that are either likely to cause stragglers or are commonly to blame in our setting. We start by looking at whether faulty hardware (or software) in one (or a few) workers is to blame (\S\ref{sec:machine-issue-root-cause}) %
Next, we discuss three common root causes in our cluster: imbalance in how computation is partitioned across pipeline stages (\S\ref{sec:fixed-group-root-cause}), imbalance in the length of sequences in each microbatch (\S\ref{sec:seqlenimba-root-cause}), and stalls during garbage collection (\S\ref{sec:gc-root-cause}). Finally, in \S\ref{sec:other-root-cause}, we briefly describe two interesting, but uncommon causes of stragglers.

\subsection{Are individual workers to blame?} 
\label{sec:machine-issue-root-cause}

We first analyze how many of the stragglers are due to hardware or software problems with workers. Because we run health-checks, we expect that only a small fraction of workers should exhibit any problems. Consequently, if a job is slow because of problems with a few workers then fixing the execution time of \segments running on the problematic workers (by changing them to the ideal execution time) should suffice to fix the entire job's completion time.\footnote{Though one faulty machine with a network issue affecting DP communication will affect additional $DP\_degree -1$ workers participating in the same DP-communication, as shown in \S\ref{sec:optype-impact} slowdown of DP communication is not a major contributor to straggling jobs, and consequently we do not consider it.} We use this observation to measure the impact of problematic workers on straggling jobs.

We start by computing the slowdown attributed to each worker, using the same technique that we used in the previous section to estimate an \segment's slowdown percent. Specifically, we define the slowdown $S_w$ for worker $w$ as
\begin{equation}
\label{eq:worker-slowdown}
    S_{w}\triangleq T_{ideal}^{-w} / T_{ideal}.
\end{equation}
where $T_{ideal}^{-w}$ is the job completion time (obtained from the simulator) when we only fix \segments that were not executed on worker $w$, and $T_{ideal}$ is the execution time when all \segments are fixed.

Next, for each job, we select the set of workers $W$ whose slowdowns $S_w$ are the highest 3\% of the job. If a small number of workers are problematic due to hardware (or software) problems, then $W$ would contain those workers. We then compute the fraction of the job's slowdown that can be attributed to these workers, $M_{W}$

\begin{equation}
M_{W}=\frac{T - T_{ideal}^{W}}{T - T_{ideal}}, 
\end{equation}
where $T$ is the simulated original step duration (without any stragglers being fixed), $T_{ideal}^W$ is the simulated step duration when only \segments running on the selected workers are fixed, and $T_{ideal}$ is the ideal simulated step duration after all stragglers are fixed.

Computing $M_{W}$ for large jobs (with thousands of workers) is expensive because it requires running thousands of simulations to compute $S_w$ for each worker $w$. Therefore, we use an approximation to scale our analysis: rather than computing a slowdown for individual workers, we measure a slowdown for whole DP ranks and PP ranks. We then assign each worker the smaller (\ie minimum) of the slowdowns for the DP and PP ranks to which it belongs (as a reminder, any worker must belong to one DP rank and one PP rank). This allows us to reduce the number of simulations from \textit{DP degree} $\times$ \textit{PP degree} to \textit{DP degree} $+$ \textit{PP degree}, making computation feasible.

\begin{figure}[t]
    \centering
    \includegraphics[width=.9\linewidth]{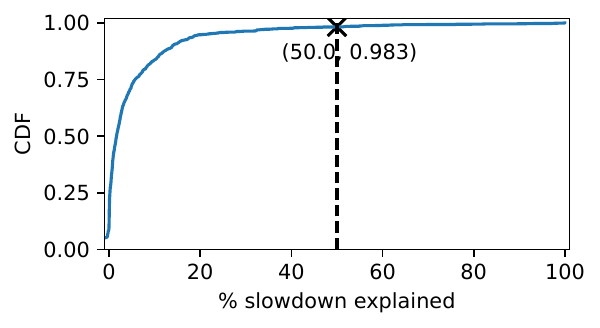}
    \caption{CDF for $M_{W}$, the performance recovered after fixing the slowest 3\% of workers, normalized by the performance recovered after fixing all workers. Observe that the slowest 3\% of workers do not contribute significantly to the observed slowdown for most jobs.}
    \label{subfig:rank-loc-v2-fixed-rank}
\end{figure}

Figure~\ref{subfig:rank-loc-v2-fixed-rank} shows a CDF of $M_{W}$ for straggling jobs. We observe that worker problems contribute to more than 50\% of the observed slowdown for only \fixedHostBlockingRatio of straggling jobs. This leads us to conclude that problematic workers are not the dominant factor for a majority of the straggling jobs in our traces.

We further investigated the few cases where problematic workers were primarily responsible for job slowdowns, and found that they resulted in a much larger slowdown: slowdown for jobs with problematic workers is \fixedHostBlockingSld compared to the average slowdown of \averageBlockingSldSJ.

\subsection{Stage Partitioning Imbalance}
\label{sec:fixed-group-root-cause}
In our traces, we observed that an imbalance between the computation performed by the last pipeline stage and other pipeline stages was a common cause of stragglers. 
The last pipeline stage needs to execute a loss layer, and in most cases, the loss layer requires more compute cycles than transformer layers (and embedding layers, which take negligible compute time)~\cite{wei2024skywork,dubey2024llama}. Thus, it is easy for users to accidentally partition the model (e.g., by evenly dividing layers over pipeline stages) in a way that leads to the last stage requiring significantly more computation, which in turn leads to stragglers.

To verify that the loss layer took longer than the transformer layers, we ran a job with four pipeline stages,
where each pipeline stage runs $9$ transformer layers, and the last stage runs an additional loss layer. The PyTorch trace for this job shows that logit computation (in the last layer) is over 9 times longer than a transformer layer, and that as a result, forward-compute (backward-compute) of the last stage is \stageImbaFwdRatioBeforeOpt (\stageImbaBwdRatioBeforeOpt) times slower than an average stage's computation.

We use a similar approach as in \S\ref{sec:machine-issue-root-cause} to study the prevalence of this problem: we compute in simulation $T_{ideal}^{lastStage}$, the job's completion time if we only fix the execution time for operations in the last pipeline stage. Next, for each job that uses pipeline parallelism, we compute $M_S$, the contribution of the last stage as $M_{S}=(T - T_{ideal}^{lastStage})/(T - T_{ideal})$ (and we set $M_S=0$ for the \noPPJobRatio of jobs that do not use pipeline parallelism).

\begin{figure}[t]
    \centering
    \includegraphics[width=.9\linewidth]{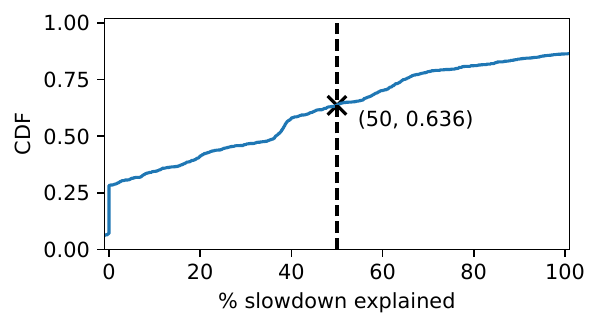}
    \caption{CDF for $M_{S}$, the performance recovered after fixing all workers on the last PP stage, normalized by the performance recovered after fixing all workers. We set $M_{S}=0$ for \noPPJobRatio of jobs not running PP. Observe that the last stage slowdown is a common straggler issue.}
    \label{subfig:rank-loc-v2-fixed-group}
\end{figure}

Figure~\ref{subfig:rank-loc-v2-fixed-group} shows the CDF of $M_S$'s across jobs. We observe that for \fixedGroupBlockingRatio of jobs, a majority of the job slowdown (\ie $M_S\geq0.5$) is because of the last pipeline stage.

Next, we looked at whether this problem could be mitigated. We already use an approach similar to Llama 3~\cite{dubey2024llama} to address this problem: we assign $\epsilon$ fewer layers to the last pipeline stage. However, we need to manually tune $\epsilon$, and this has proven to be challenging for several reasons: First, when partitioning the model, we must assign an entire transformer layer to a pipeline stage, which  limits how computation is split between pipeline stages; 
Second, the time taken by the loss layer (in proportion to the transformer layer) increases either if the vocabulary size grows, or if the maximum-sequence-length or the hidden layer size decreases. Larger ratios are likely to become more common as vocabulary size increases~\cite{tao2024scaling}, and consequently, as others have also observed concurrently~\cite{yeung2024balancing}, this is going to increase the number of transformer layers that need to be executed in earlier pipeline stages to ensure that they take the same time as last stage.
This in turn limits the number of pipeline stages that can be used for these models and leads to more layers in each virtual stage than desirable, limiting model parallelism.
Consequently, even a good value of $\epsilon$ can result in suboptimal performance.

We also attempted to manually tune the number of layers in each pipeline stage for the job (with $9$ transformer layers) that we discussed above, and we found that with manual partitioning we could get a speedup of \stageImbaImprove. However, even with manual tuning, the computational load is not perfectly even across stages, \eg the forward-compute of the last stage is \stageImbaFwdRatioAfterOpt of the other stages after tuning.

\subsection{Sequence Length Imbalance} 
\label{sec:seqlenimba-root-cause}

\begin{figure}
    \centering
    \includegraphics[width=\linewidth]{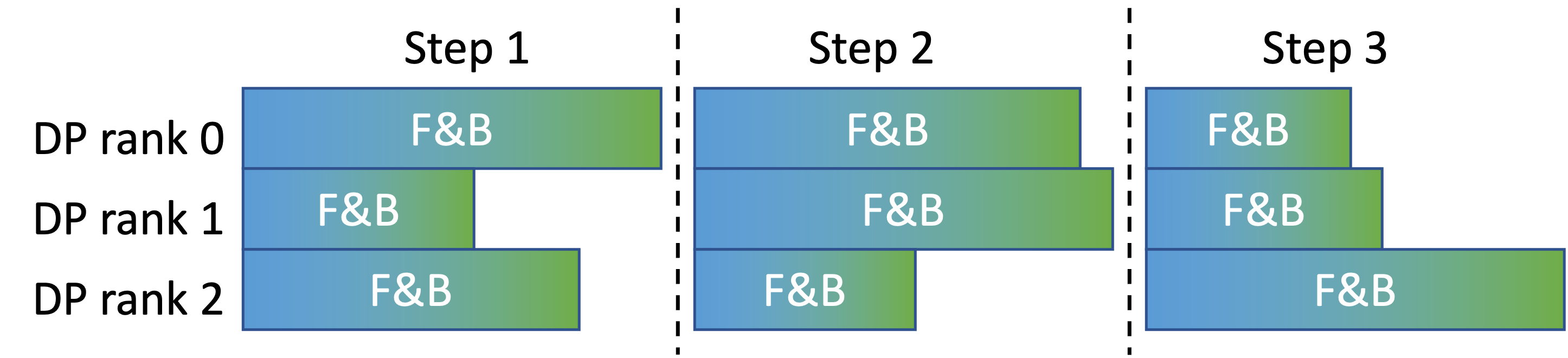}
    \caption{A representative timeline for jobs with sequence length variance and using pure data-parallelism. 
    Large variance exists in the computation \segment times, causing stragglers to happen on some random DP rank(s) every step. ``F\&B'' blocks denote the duration from when the first forward-compute launches to when the last backward-compute finishes. 
    }
    \label{fig:seqlen-imba-timeline}
\end{figure}

\begin{figure}[t]
    \centering
    \includegraphics[width=.9\linewidth]{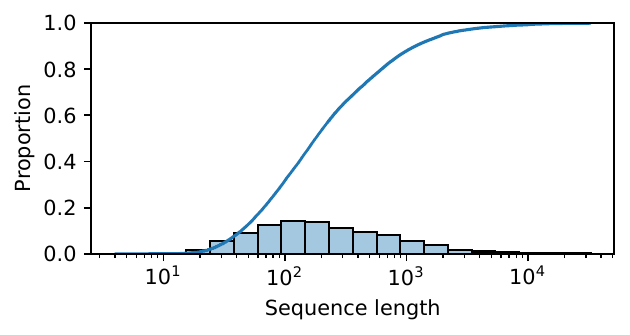}
    \caption{Sequence length distribution, collected on a job with a maximum-sequence-length of 32K. The histogram is shown in bars, and the CDF is shown on the curve. Note the log scale on the $x$-axis.}
    \label{fig:seqlen-dist}
\end{figure}

Next, we analyzed jobs whose slowdown is not explained by slow workers (\S\ref{sec:machine-issue-root-cause}) or imbalance in the computational load between pipeline stages (\S\ref{sec:fixed-group-root-cause}). Our inspection revealed that for long-context jobs, differences in sequence lengths between training data are a significant contributor to job slowdown. Recent studies~\cite{bai2024longalign,jiang2023dynapipe,ge2024enabling} have shown that the datasets used to train long-context LLMs have a long-tailed sequence length distribution. We confirmed that this was also the case in our cluster: 
Figure~\ref{fig:seqlen-dist} shows the sequence length distribution for the training data used in a long-context training job, where the \emph{maximum-sequence-length} is set to 32K.

Variance in sequence length is a problem because the algorithmic complexity of the self-attention layer is quadratic~\cite{korthikanti2023reducing}. Our system forms a training microbatch by collecting sequences (chosen at random) until the total length of the microbatch reaches a predefined maximum-sequence-length. However, the computation time for the microbatch depends not just on the maximum-sequence-length, but also on the length of each sequence $s_i$ that is included in the microbatch because the computation time for a microbatch is $O(\sum{s_i^2})$. For instance, a microbatch with one sequence of length 32K requires $32\times$ more compute time than a microbatch containing 32 sequences of length 1K. The difference in computation time across microbatches results in bubbles in pipeline stages, and in 
DP ranks finishing at different times (shown in Figure~\ref{fig:seqlen-imba-timeline}). Both combine to produce stragglers.

\begin{figure}[t]
    \centering
    \includegraphics[width=\linewidth]{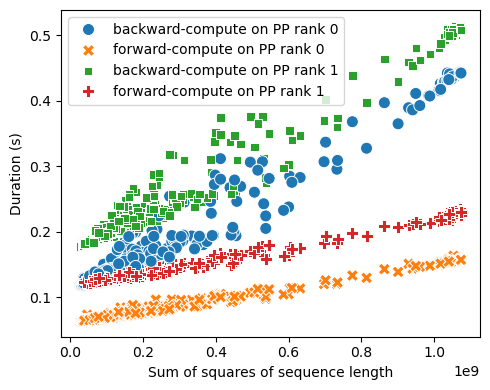}
    \caption{Microbatch computation duration vs sum of sequence length squares, profiled over dozens of training steps on a job with maximum-sequence-length of 32K.
    Each data point represents a microbatch on the forward or backward pass on a specific PP rank.
    }
    \label{fig:time-sos}
\end{figure}

Our reasoning above is based on the observation that computation time is proportional to $O(\sum{s_i^2})$. In Figure~\ref{fig:time-sos}, we empirically verify this assumption: we plot the duration and $\sum_i s_i^2$ of each microbatch in the first several training steps of a representative job, and find that indeed they are proportional.

\begin{figure}[t]
    \centering
    \includegraphics[width=.8\linewidth]{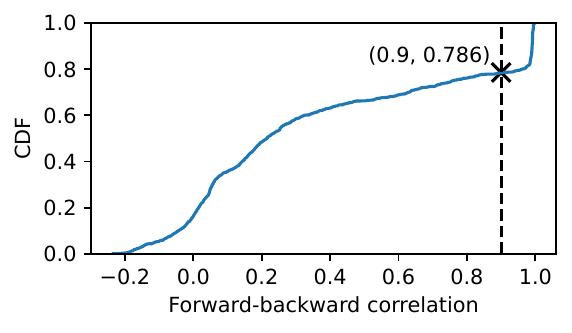}
    \caption{CDF of forward-backward correlation of all straggling jobs in the solid line. The dashed line shows the proportion of overall resource waste by the jobs with correlation under the $x$ values.}
    \label{fig:corr-cdf-all}
\end{figure}

Next, we quantified the fraction of jobs whose slowdown is due to sequence length imbalance. We cannot use the same metric as the previous two sections because our trace does not provide us with enough information to correct for sequence length imbalance. Instead, we measure the occurrence of slowdown due to sequence length imbalance using a \emph{forward-backward correlation} metric. Our metric is based on the observation that if forward-compute for a microbatch is slow because of sequence length imbalance, then the backward-compute must also slow down by a similar amount, and the forward- and backward-compute times must be correlated.  
Figure~\ref{fig:corr-cdf-all} shows the Pearson correlation factor for a pipeline stage\footnote{To avoid noise brought by microbatches computing loss and embedding layers, we only use microbatches on the second PP stage when \textit{PP degree} $\geq 3$. Otherwise, we use the first stage, and if VPP is used we also drop the first virtual stage to discard microbatches containing embedding layers.} in each straggling job (i.e., those with $S\geq 1.1$). Empirically, we found that jobs with a correlation coefficient $\geq 0.9$ were most likely to have been slowed down because of sequence length imbalance. Using this threshold, we observe that \seqlenimbaJobRatio jobs are affected by sequence length imbalance. They have an average slowdown of \seqlenimbaSlowdown.

\begin{figure}[t]
    \centering
    \includegraphics[width=\linewidth]{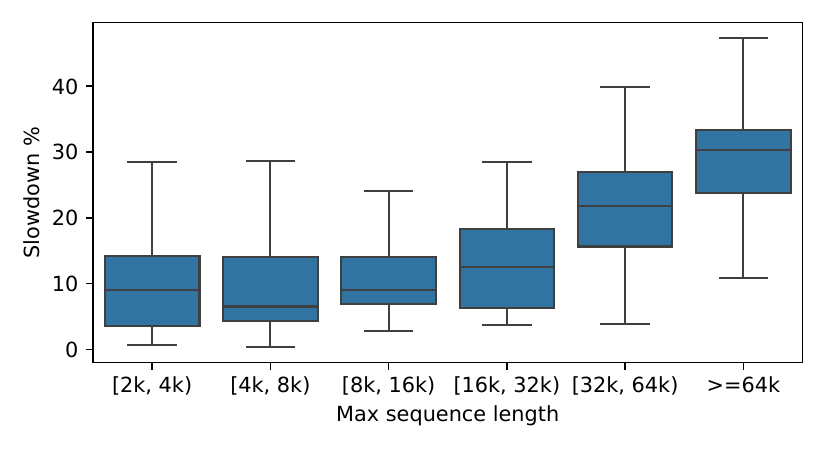}
    \caption{Long context vs others.}
    \label{fig:lctx}
\end{figure}

Furthermore, in Figure~\ref{fig:lctx} we analyzed how a change in the maximum-sequence-length affects job slowdown. We observed that sequence length imbalance has a larger effect as the maximum-sequence-length grows. But context lengths are growing, and thus addressing this scalability challenge is becoming increasingly important.

One approach to addressing this problem is to distribute sequences across microbatches so as to equalize the computation time for all microbatches. As we discussed above, microbatch computation times can be accurately predicted given sequence lengths (Figure~\ref{fig:time-sos}). Thus, we prototyped a version of this approach where after a batch is formed, we use a linear model to redistribute sequences so that all DP ranks have a balanced computational load. We formulated the redistribution problem as a multiway number partitioning problem~\cite{wikimpp} and solved it using a greedy algorithm similar to the one used by DistTrain~\cite{zhang2024disttrainaddressingmodeldata}.\footnote{Except that we sort sequences in descending order since it gives a much better result in our case.} We used PyTorch's distributed KVStore~\cite{torchtcpstore} to exchange sequences between DP ranks. Next, each DP rank divides the assigned sequences into microbatches so that the sums of sequence lengths across microbatches are balanced, and we achieve this with a greedy algorithm. We tested this approach on a representative job with a maximum-sequence-length of $32$K,
and observed a \seqlenimbaExpSpeedup improvement in throughput.

Deploying such a fix in practice requires evaluating its effect at scale, and we leave this to future work. We expect that our proposed fix might increase memory requirements, because balancing computation in this manner results in sequence length sums varying across DP ranks, and might lead to increased memory requirements for some ranks. Furthermore, this fix only solves the imbalance at the DP level, but similar to DistTrain and DynaPipe~\cite{jiang2023dynapipe}, we also observed imbalance at the PP level (and a large number of PP bubbles) in jobs with large PP degrees and more microbatches. Other approaches to balancing computation across microbatches might be required to address PP-level imbalance.

\subsection{Python's Automatic Garbage Collection}
\label{sec:gc-root-cause}
\begin{figure}[t]
    \centering
    \includegraphics[width=\linewidth]{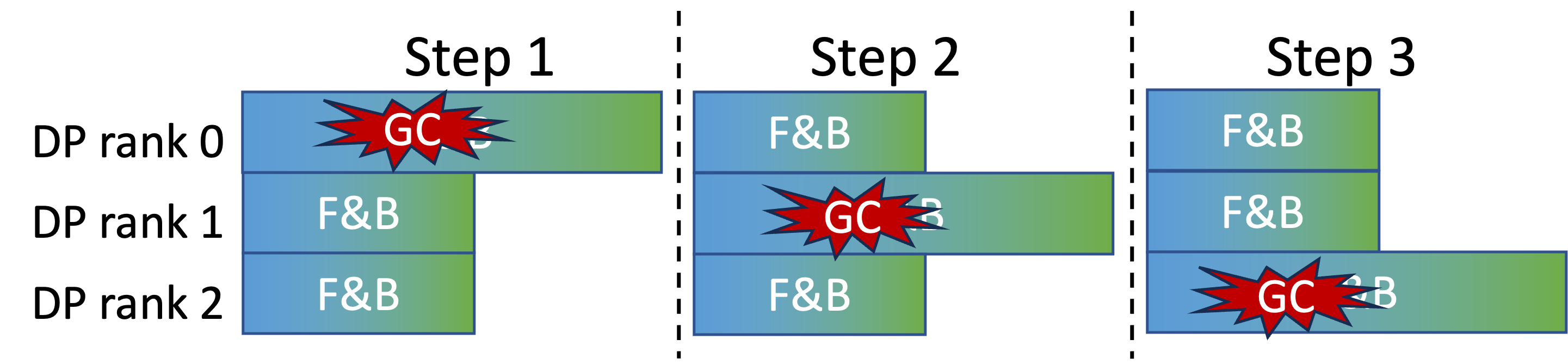}
    \caption{Representative timeline for jobs suffering from the GC straggler. Different workers perform their GC at different steps, straggling each other.}
    \label{fig:gc-timeline}
\end{figure}
Garbage collection (GC) is another significant cause of slowdowns. The jobs in our cluster are run using Python, whose runtime triggers GC when it is deemed necessary~\cite{cpythongc}. Once triggered, GC can take 100s of milliseconds, and during this period the user program is paused and new kernels cannot be started,\footnote{Python uses a stop-the-world garbage collector.} stalling forward-compute \segments. Note that backward-compute \segments are not affected because they are launched from C++.

This problem is further exacerbated because different Python processes trigger GC at different times. While the GC pauses a single worker, it blocks the entire training job
as shown in Figure~\ref{fig:gc-timeline}. The number of such pauses grows as the number of workers increases, and thus stragglers due to GC have become a more important concern as models have increased in scale.

Therefore, in 2023 our engineering team implemented a \gcOpt optimization to mitigate stragglers due to GCs. This optimization turns off Python's automatic GC mechanism, and instead manually schedules GC at a user-specified interval (that is specified in terms of training steps). This ensures that all workers run GC at the same time. This optimization is effective at preventing GC-related stragglers: on a job that uses 128 DP ranks, using this optimization so that GC is run every 500 training steps results in a \gcExpSpeedup improvement.

While the \gcOpt optimization is effective, it is challenging to use in practice because choosing an appropriate GC-interval is hard. Picking too large a GC-interval can lead to the job running out of memory and crashing, while choosing too small a GC-interval can lead to worse performance. But the rate at which memory is allocated, and thus the appropriate GC-interval varies by job, and thus users need to tune the interval for each job. Therefore, we are conservative and do not enable \gcOpt by default: users can turn it on after they have analyzed previous runs of a job to determine an appropriate GC-interval. Consequently, this fix has not been widely adopted in our cluster.

We also observed that the amount of time for which the GC pauses execution increases as the job progresses. This leads to a gradual decrease in training throughput, as has also been observed in the past~\cite{jiang2024megascale, imbue}.
We suspect the slowdown is due to a memory leak, which increases heap size as the job runs longer and leads to longer GC pauses.
Interestingly, we found that \gcOpt can mask the impact of this leak and leads to sustained training throughput.

\subsection{Other Root Causes}
\label{sec:other-root-cause}
As we stated previously, the set of root causes we have discussed is not exhaustive, and jobs in our traces might have been slowed down by other problems. Rather, our goal was to analyze some of the more obvious or common causes of stragglers. Below, we describe two uncommon but interesting root causes that appeared in our traces.

\paragraph{CUDA memory fragmentation.}
In a few cases, we observed that memory fragmentation slowed down PyTorch's CUDA memory allocator, and led to a significant increase in calls to \texttt{cudaFree} and \texttt{cudaMalloc}, which led to slower than usual forward- or backward-compute \segments. In our traces, this appears as cases where TP communication kernels running on different TP ranks in the same TP group start at different times but finish roughly at the same time, suggesting that some of the communication kernels experience launch delays. We examined the PyTorch traces for these jobs to confirm that they did indeed make several \texttt{cudaFree} and \texttt{cudaMalloc} calls, and then we enabled allocator memory tracing and reran the job to further understand the problem. During rerun we observed many \texttt{segment\_alloc} and \texttt{segment\_free} calls, confirming that the PyTorch memory allocator was to blame. In future work, we plan to study both the prevalence of this problem and mitigation strategies.

\paragraph{False kernel dependency.}
When we first started training large MoE models, we found that reduce-scatter kernels (used to synchronize gradients) block other kernels (that do not depend on the reduce-scatter kernel) from being launched, and lead to significant job slowdown. We suspect that this is because of false dependencies~\cite{ng2023paella,falsedependency} caused by unrelated kernels sharing the same CUDA hardware queue. We found that increasing \texttt{CUDA\_DEVICE\_MAX\_CONNECTIONS} alleviates the problem. However, we have also found that the problem seems to occur (and then disappear) as we evolve our model and framework, and we continue to investigate it to identify the cause.

\subsection{Summary: Observations and Implications}
\label{sec:new-obs}
\newtext{
Our root cause analysis yields the following key observations about our traces:
\begin{itemize}
    \item Relatively few stragglers were due to machine problems (either due to hardware or software), implying that traditional health metrics are unlikely to aid in detecting or preventing most stragglers.
    \item The most prevalent causes for straggling are due to uneven pipeline stage partitioning, imbalance in sequence lengths in each microbatch, and pauses induced by the garbage collector.  Our observations also led us to develop a new approach to address sequence length imbalance (\S\ref{sec:seqlenimba-root-cause}).
    \item We observed two new but less prevalent causes for stragglers: PyTorch memory fragmentation and false kernel dependencies. While both issues had been reported before as potential performance problems, they were not known to lead to stragglers prior to our work.
\end{itemize}
}

\section{Validation of Simulation Fidelity}\label{sec:validate}
Our analysis uses a simulator to estimate a job's execution time in the absence of stragglers, and also after fixing some stragglers. 
\newtext{We validate the simulator's accuracy in two ways: 1) by comparing a job's average step time in the simulated original timeline to its average step time in the actual timeline; 2) by artificially injecting stragglers and comparing simulated and actual slowdowns.}

\paragraph{Simulation inaccuracy and its sources.} \newtext{
As we described in \S\ref{sec:method}, not all types of \segments are recorded in the trace: specifically, \segments run on the CPU, including data loading, are omitted from the trace.
In most cases, this limitation does not pose a problem because these CPU \segments are overlapped with GPU ones, allowing their latency to be hidden. However, the overlap is not always perfect and delays the launch of the next \segment. This \emph{launch delay} is not simulated, constituting the main source of discrepancy in the simulation.
To measure the discrepancy, we compute the average step run-time $\tau=T/n$ of the simulated original timeline (\S\ref{sec:gen:what-if}) for a job with $n$ steps and compare it with the actual step run-time $\tau_{act}$.
In our experiments, the simulation discrepancy has a median of 1.3\% and 90-percentile of 5.5\%. }

After examining the traces with large simulation discrepancies, we found three main causes: (a) As data is stored on a remote storage cluster using a separate, slower network, data loading is susceptible to network slowdown and time-out, leading to long launch delays for the first forward-compute operations of some training steps; (b) Long context jobs can have long delays before the launch of the first forward-compute operations of each step because, as part of batch preparation, samples are padded to the maximum sequence length, which is time-consuming; and (c) Early deployments of the \gcOpt optimization, where the GC was run every few training steps before gradient synchronization, slowed down the launch of those \segments. We have resolved the latter two problems, but all three affect some of the traces that we captured. To ensure analysis fidelity, we dropped any traces where the simulation discrepancy is larger than 5\%.

\paragraph{Validating the Accuracy of Slowdown Estimation.} We validate the accuracy of slowdown estimation by running a job whose DP, PP, and TP degrees were 4, and \cpname degree was 1. We artificially slow down the first rank, \ie the worker with global-rank 0, by running a background process that periodically performs multiple matrix multiplications (MatMuls) of size $10 \text{K}\times 10 \text{K}$. By varying the time interval between MatMul launches, we create three levels of slowdown and measure the resulting slowdown. We also used the simulator to compute a simulated slowdown. Our measured slowdowns were \mildAct, \mediumAct, and \severeAct, respectively, which are close to the estimated slowdowns \mildEst, \mediumEst, and \severeEst.

\section{Limitation}\label{s:limitations}
Next, we discuss the limitations of our analysis. Our limitations are due to the trace contents captured by \ndtimer, and we plan to address these limitations in future work. 

\paragraph{Limitations due to data captured by \ndtimer.} \ndtimer performs coarse-grained profiling and records time taken by a microbatch's forward and backward computation. It is challenging to use this information to analyze stragglers that occur within a TP or \cpname group, because TP and \cpname groups synchronize, and stragglers at the TP or \cpname granularity show up as slow microbatches in our traces. As a result, if the straggler slows down all microbatches, our method of estimating idealized microbatch durations will not allow us to compute the ideal straggler-free execution time, and thus we cannot analyze the effect of such stragglers.

A second limitation is due to a bug in \ndtimer which affects some 
of our traces. This bug caused \ndtimer to not record some \segments, resulting in simulating the launch of some forward- and backward-compute \segments earlier incorrectly. The bug was spotted after we had begun collecting traces, and we post-processed traces affected by this bug to fix this problem.

\paragraph{Job covered by our analysis.}
\label{sec:trace-coverage}
To ensure analysis fidelity, we had to discard a portion of the traces we collected from our cluster. \newtext{Consequently, our analysis does not cover all the LLM training jobs run within the cluster,
and may underestimate straggler prevalence and severity.}
\newtext{The reasons for discarding are mostly due to problems with the trace and we outline them below.}

First, we eliminated any jobs that repeatedly fail. Our cluster uses a system that automatically resubmits jobs that fail. However, software bugs result in cases where a single job might fail dozens of times or more, and we eliminate these jobs to avoid introducing bias into our analysis. In particular, we discard any job that is restarted more than 15 times, which results in us discarding $\manyRunsJobRatio$ of the recorded jobs or $\manyRunsGPUHourRatio$ of the recorded GPU hours.

Second, we discarded jobs for which we could not successfully run the what-if analysis. This results in us discarding $\simJobRatio$ of the remaining jobs, which amounts to $\simGPUHourRatio$ of the remaining GPU hours. 
\newtext{Most of these are due to issues with the trace:}
28\% of the traces were discarded because we could not parse the job's command line to determine the degree of parallelism; 28\% were discarded because the job had too few steps to allow analysis;\footnote{We filter out steps that are doing batch size warmup, and in some cases this results in jobs with insufficient steps.} and 25\% were due to corrupt traces. 

Finally, as noted earlier (\S\ref{sec:gen:what-if}), we discard jobs with large simulation discrepancy ($>5\%$), which amounts to $\largeSimErrorJobRatio$ and $\largeSimErrorGPUHourRatio$ of the remaining jobs and GPU hours, respectively.
This results in total coverage of $\totCovJobRatio$ jobs and $\totCovGPUHourRatio$ GPU hours.

\section{Online Straggler Detection and Diagnostics with What-if Analysis}
\label{sec:deployment}

\begin{figure}[t!]
    \centering

\subcaptionbox{Worker issue\label{subfig:heatmap-fixed-host}}{%
  \includegraphics[width=\linewidth]{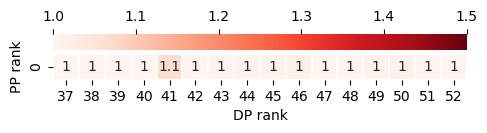}%
}
\\
\subcaptionbox{Stage partitioning imbalance\label{subfig:heatmap-fixed-group}}{%
  \includegraphics%
  [width=\linewidth]{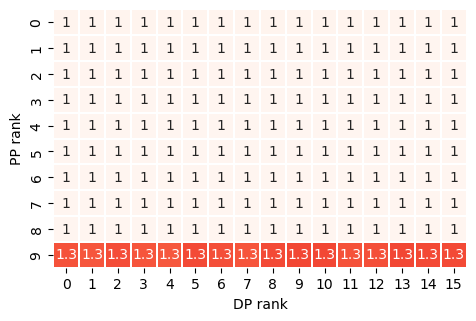}%
}
\\
\subcaptionbox{Sequence length imbalance
\label{subfig:heatmap-uniform}}{%
  \includegraphics[width=0.6\linewidth]{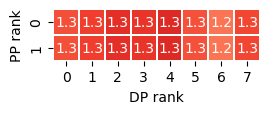}%
}
    \caption{Heatmap patterns with different root causes.}
    \label{fig:heatmap}
\end{figure}

To make it easier for users to benefit from the analysis we have discussed so far, we built an online monitoring service called \emph{\sysnameNoXSpace} that runs automatically after each \ndtimer profiling session (which records dozens of training steps). \sysname
estimates slowdown, per-step slowdown, and worker slowdown, and presents the results on a webpage.
Similar to Pingmesh~\cite{guo2015pingmesh}, we present worker slowdowns using a heatmap,
where each cell represents a worker with its x- and y-coordinate as its DP and PP rank, respectively, and the color depth represents the worker's slowdown.
Our use of such a heatmap serves two purposes: first, it makes finding straggling workers easier, and second, the pattern of slowdowns often helps pinpoint the initial root cause for the slowdown, making it easier to address the problem.
For instance, jobs with worker issues, stage partitioning imbalance and sequence length variance have distinct patterns as shown in Figure~\ref{fig:heatmap}.
\sysname also presents a per-step heatmap using per-step duration instead of average in Equation~(\ref{eq:worker-slowdown}) when computing worker slowdowns, to reflect only the straggling within each step.

We have configured \sysname to alert our on-call team whenever important jobs experience significant slowdowns.
When alerted, the on-call team checks
the worker heatmap to identify a suspected root cause by seeing if it matches one of the known patterns.
We then use the per-step slowdown and the per-step heatmap to locate the problematic step and ranks to further drill down and understand the problem.%

Within a month of its deployment, \sysname has helped us identify and address several stragglers: it allowed us to locate faulty machines in three cases where machine problems were responsible for stragglers, it helped identify a case where sequence length variance was leading to large job slowdown, and a separate case where imbalanced compute partitioning across pipeline stages was to blame.

\section{Related Work}
\paragraph{Straggler in big data.}
The straggler problem has been extensively studied in big data frameworks such as MapReduce~\cite{ananthanarayanan2010reining,ousterhout2015making,gill2020tails,mapreduce}, predating the era of deep learning.
Mantri~\cite{ananthanarayanan2010reining} provided an empirical characterization of stragglers in large-scale MapReduce jobs, where they quantify the straggler prevalence and attribute the root causes to load-imbalance, cross-rack traffic, and bad machines.
They propose a system that actively restarts or duplicates outlier tasks while optimizing task placement and scheduling to mitigate delays. 
Ousterhout et al\cite{ousterhout2015making} conducted a more in-depth analysis of the performance bottleneck of Spark using a what-if simulation similar to our method. They attribute stragglers to diverse causes, including Java GC and file system delays.

\newtext{However, the simulation approaches and findings of~\cite{ananthanarayanan2010reining, ousterhout2015making} are not applicable to LLM training. First, much of the complexity for simulating MapReduce jobs is due to dynamic task scheduling, \eg a reduce task is scheduled to a machine after its inputs become available at runtime. In contrast, LLM training uses simple static scheduling, \ie tasks are placed at the beginning of the job and never changed. 
Second, the complexity of simulating LLM training jobs lies in handling the complex dependency structure stemming from mixed parallelism. In particular, LLM jobs have to account for pipeline parallelism, tensor parallelism, and data parallelism, each of which occurs at different granularities and can involve only a subset of ranks. By contrast, dependencies of MapReduce jobs are much simpler, \ie all map tasks communicate with all reduce tasks. %
These differences limit the direct applicability of their approaches and conclusions to LLM training.
}

\paragraph{Straggler in deep learning training.}
Straggler problems have also been studied in the context of data-parallelism in the deep learning era.
Many of these prior proposals have focused on developing straggler mitigation strategies: Ramamoorthy et al~\cite{ramamoorthy2020straggler} suggest using redundant coded computations to mitigate stragglers; Downpour SGD~\cite{dean2012large} and Project Adam~\cite{chilimbi2014project} use asynchronous SGD and stale gradient updates to mitigate stragglers; Hop~\cite{luo2019hop}, 
Tensorflow~\cite{tensorflow} and Chen et al~\cite{chen2016revisiting} 
use backup workers to resolve stragglers; and DropCompute~\cite{giladi2023dropcompute} suggests dropping updates from stragglers. By contrast, our work focuses on characterizing the effect of stragglers and their root cause.

\paragraph{Straggler in LLM training.}
Recent studies have begun to explore stragglers in LLM training.
Malleus~\cite{li2024malleus} mitigates the impact of stragglers using dynamic parallelism adjustments based on real-time device performance.
MegaScale~\cite{jiang2024megascale}, Llama3~\cite{dubey2024llama} and Imbue's report~\cite{imbue} discuss infrastructure challenges for LLM training and briefly mention how they debug stragglers, but do not provide in-depth analyses.
Unlike these works, we focus on a detailed and comprehensive analysis of the straggler problem itself.

\falcon~\cite{wu2024falcon} presents a detailed characterization of stragglers and introduces several mitigation strategies. 
\newtext{
Our work differs in several ways. 
First, we analyze jobs on a dedicated cluster while \falcon's analysis is for a shared cluster. The difference in the settings lead to different results: unlike \falcon, we did not see stragglers due to resource contention. 
Second, \falcon's analysis of large jobs ($512$ to $1024$ GPUs) is only limited to 27 job traces, while our analysis has a significantly larger scale: among the $\totCovJobCnt$ traces we analyzed, 562 jobs use $\ge 512$ GPUs. 
With only a few dozen traces, \falcon relies on manual analysis to study stragglers and determine their root cause. By contrast, we use a semi-automated approach combining simulation-based what-if analysis and manual verification of hypothesized root causes. Furthermore, \falcon's analysis overlooks stragglers that affect most steps rather than only a small fraction of steps. As we show in \S\ref{subsec:steps}, the former is more common than the latter in our traces.

}

\section{Conclusion}
We present an in-depth study of the straggler problem in LLM training on job traces collected from our cluster. 
We use what-if analysis as our core approach, which allows us to estimate slowdown caused by stragglers as well as attribute it to different workers or \segment types.
These measurements help us study the overall impact of stragglers, characterize straggler symptoms from various aspects, and preliminarily diagnose the root cause.
Through extensive case studies, we find that stage partitioning imbalance, sequence length imbalance, and Python's automatic GC are the main root causes; machine issues are rarely observed to cause stragglers, but they tend to cause severe slowdown when happening.

We have implemented parts of the analysis pipeline in \sysname, a monitoring tool that is deployed in the training cluster at \company. \sysname allows us to more accurately detect stragglers in running jobs and improves the diagnostics efficiency.%

\section*{Acknowledgement}
\newtext{We thank OSDI reviewers and our shepherd for their insightful feedback. We also thank Lingfan Yu, Zhihao Zhang, Tao Wang and Lei Zhang for insightful discussions. Haibin Lin and Jinyang Li are the corresponding authors.}

\clearpage

\bibliographystyle{plainnat}
\bibliography{reference}

\begin{thebibliography}{43}
\providecommand{\natexlab}[1]{#1}
\providecommand{\url}[1]{\texttt{#1}}
\expandafter\ifx\csname urlstyle\endcsname\relax
  \providecommand{\doi}[1]{doi: #1}\else
  \providecommand{\doi}{doi: \begingroup \urlstyle{rm}\Url}\fi

\bibitem[Abadi et~al.(2016)Abadi, Barham, Chen, Chen, Davis, Dean, Devin, Ghemawat, Irving, Isard, et~al.]{tensorflow}
Mart{\'\i}n Abadi, Paul Barham, Jianmin Chen, Zhifeng Chen, Andy Davis, Jeffrey Dean, Matthieu Devin, Sanjay Ghemawat, Geoffrey Irving, Michael Isard, et~al.
\newblock Tensorflow: A system for large-scale machine learning.
\newblock In \emph{12th {USENIX} symposium on operating systems design and implementation ({OSDI} 16)}, pages 265--283, 2016.

\bibitem[Ananthanarayanan et~al.(2010)Ananthanarayanan, Kandula, Greenberg, Stoica, Lu, Saha, and Harris]{ananthanarayanan2010reining}
Ganesh Ananthanarayanan, Srikanth Kandula, Albert Greenberg, Ion Stoica, Yi~Lu, Bikas Saha, and Edward Harris.
\newblock Reining in the outliers in $\{$Map-Reduce$\}$ clusters using mantri.
\newblock In \emph{9th USENIX Symposium on Operating Systems Design and Implementation (OSDI 10)}, 2010.

\bibitem[Ananthanarayanan et~al.(2013)Ananthanarayanan, Ghodsi, Shenker, and Stoica]{ananthanarayanan2013effective}
Ganesh Ananthanarayanan, Ali Ghodsi, Scott Shenker, and Ion Stoica.
\newblock Effective straggler mitigation: Attack of the clones.
\newblock In \emph{10th USENIX Symposium on Networked Systems Design and Implementation (NSDI 13)}, pages 185--198, 2013.

\bibitem[Bai et~al.(2024)Bai, Lv, Zhang, He, Qi, Hou, Tang, Dong, and Li]{bai2024longalign}
Yushi Bai, Xin Lv, Jiajie Zhang, Yuze He, Ji~Qi, Lei Hou, Jie Tang, Yuxiao Dong, and Juanzi Li.
\newblock Longalign: A recipe for long context alignment of large language models.
\newblock \emph{arXiv preprint arXiv:2401.18058}, 2024.

\bibitem[Bhatnagar et~al.(2023)Bhatnagar, Coutinho, Feng, Liu, Lin, Feng, Acar, and Huang]{hta}
Anupam Bhatnagar, Brian Coutinho, Xizhou Feng, Yifan Liu, Sung-Han Lin, Louis Feng, Michael Acar, and Yuzhen Huang.
\newblock Holistic trace analysis.
\newblock \url{https://hta.readthedocs.io/en/latest/index.html}, 2023.

\bibitem[Chen et~al.(2016)Chen, Pan, Monga, Bengio, and Jozefowicz]{chen2016revisiting}
Jianmin Chen, Xinghao Pan, Rajat Monga, Samy Bengio, and Rafal Jozefowicz.
\newblock Revisiting distributed synchronous sgd.
\newblock \emph{arXiv preprint arXiv:1604.00981}, 2016.

\bibitem[Chilimbi et~al.(2014)Chilimbi, Suzue, Apacible, and Kalyanaraman]{chilimbi2014project}
Trishul Chilimbi, Yutaka Suzue, Johnson Apacible, and Karthik Kalyanaraman.
\newblock Project adam: Building an efficient and scalable deep learning training system.
\newblock In \emph{11th USENIX symposium on operating systems design and implementation (OSDI 14)}, pages 571--582, 2014.

\bibitem[Curtsinger and Berger(2015)]{curtsinger2015coz}
Charlie Curtsinger and Emery~D Berger.
\newblock Coz: Finding code that counts with causal profiling.
\newblock In \emph{Proceedings of the 25th Symposium on Operating Systems Principles}, pages 184--197, 2015.

\bibitem[Dean and Ghemawat(2004)]{mapreduce}
Jeffrey Dean and Sanjay Ghemawat.
\newblock {MapReduce}: Simplified data processing on large clusters.
\newblock In \emph{6th Symposium on Operating Systems Design \& Implementation (OSDI 04)}, San Francisco, CA, December 2004. USENIX Association.
\newblock URL \url{https://www.usenix.org/conference/osdi-04/mapreduce-simplified-data-processing-large-clusters}.

\bibitem[Dean et~al.(2012)Dean, Corrado, Monga, Chen, Devin, Mao, Ranzato, Senior, Tucker, Yang, et~al.]{dean2012large}
Jeffrey Dean, Greg Corrado, Rajat Monga, Kai Chen, Matthieu Devin, Mark Mao, Marc'aurelio Ranzato, Andrew Senior, Paul Tucker, Ke~Yang, et~al.
\newblock Large scale distributed deep networks.
\newblock \emph{Advances in neural information processing systems}, 25, 2012.

\bibitem[Dubey et~al.(2024)Dubey, Jauhri, Pandey, Kadian, Al-Dahle, Letman, Mathur, Schelten, Yang, Fan, et~al.]{dubey2024llama}
Abhimanyu Dubey, Abhinav Jauhri, Abhinav Pandey, Abhishek Kadian, Ahmad Al-Dahle, Aiesha Letman, Akhil Mathur, Alan Schelten, Amy Yang, Angela Fan, et~al.
\newblock The llama 3 herd of models.
\newblock \emph{arXiv preprint arXiv:2407.21783}, 2024.

\bibitem[Fan et~al.(2021)Fan, Rong, Meng, Cao, Wang, Zheng, Wu, Long, Yang, Xia, et~al.]{fan2021dapple}
Shiqing Fan, Yi~Rong, Chen Meng, Zongyan Cao, Siyu Wang, Zhen Zheng, Chuan Wu, Guoping Long, Jun Yang, Lixue Xia, et~al.
\newblock Dapple: A pipelined data parallel approach for training large models.
\newblock In \emph{Proceedings of the 26th ACM SIGPLAN Symposium on Principles and Practice of Parallel Programming}, pages 431--445, 2021.

\bibitem[Forum(2022)]{falsedependency}
NVIDIA Forum.
\newblock How many streams? maximum number of streams, 2022.

\bibitem[Ge et~al.(2024)Ge, Fu, Li, Wang, Lin, Wang, Nie, Zhang, Miao, and Cui]{ge2024enabling}
Hao Ge, Fangcheng Fu, Haoyang Li, Xuanyu Wang, Sheng Lin, Yujie Wang, Xiaonan Nie, Hailin Zhang, Xupeng Miao, and Bin Cui.
\newblock Enabling parallelism hot switching for efficient training of large language models.
\newblock In \emph{Proceedings of the ACM SIGOPS 30th Symposium on Operating Systems Principles}, pages 178--194, 2024.

\bibitem[Giladi et~al.(2023)Giladi, Gottlieb, Shkolnik, Karnieli, Banner, Hoffer, Levy, and Soudry]{giladi2023dropcompute}
Niv Giladi, Shahar Gottlieb, Moran Shkolnik, Asaf Karnieli, Ron Banner, Elad Hoffer, Kfir~Yehuda Levy, and Daniel Soudry.
\newblock Dropcompute: simple and more robust distributed synchronous training via compute variance reduction.
\newblock \emph{arXiv preprint arXiv:2306.10598}, 2023.

\bibitem[Gill et~al.(2020)Gill, Ouyang, and Garraghan]{gill2020tails}
Sukhpal~Singh Gill, Xue Ouyang, and Peter Garraghan.
\newblock Tails in the cloud: a survey and taxonomy of straggler management within large-scale cloud data centres.
\newblock \emph{The Journal of Supercomputing}, 76:\penalty0 10050--10089, 2020.

\bibitem[Guo et~al.(2015)Guo, Yuan, Xiang, Dang, Huang, Maltz, Liu, Wang, Pang, Chen, et~al.]{guo2015pingmesh}
Chuanxiong Guo, Lihua Yuan, Dong Xiang, Yingnong Dang, Ray Huang, Dave Maltz, Zhaoyi Liu, Vin Wang, Bin Pang, Hua Chen, et~al.
\newblock Pingmesh: A large-scale system for data center network latency measurement and analysis.
\newblock In \emph{Proceedings of the 2015 ACM Conference on Special Interest Group on Data Communication}, pages 139--152, 2015.

\bibitem[Huang et~al.(2019)Huang, Cheng, Bapna, Firat, Chen, Chen, Lee, Ngiam, Le, Wu, et~al.]{huang2019gpipe}
Yanping Huang, Youlong Cheng, Ankur Bapna, Orhan Firat, Dehao Chen, Mia Chen, HyoukJoong Lee, Jiquan Ngiam, Quoc~V Le, Yonghui Wu, et~al.
\newblock Gpipe: Efficient training of giant neural networks using pipeline parallelism.
\newblock \emph{Advances in neural information processing systems}, 32, 2019.

\bibitem[Inc.()]{ndtimeline}
ByteDance Inc.
\newblock Ndtimeline --- vescale.
\newblock \url{"https://github.com/volcengine/veScale/blob/main/vescale/ndtimeline/README.md"}.

\bibitem[Jiang et~al.(2023)Jiang, Jia, Zheng, Wang, and Wu]{jiang2023dynapipe}
Chenyu Jiang, Zhen Jia, Shuai Zheng, Yida Wang, and Chuan Wu.
\newblock Dynapipe: Optimizing multi-task training through dynamic pipelines.
\newblock \emph{arXiv preprint arXiv:2311.10418}, 2023.

\bibitem[Jiang et~al.(2024)Jiang, Lin, Zhong, Huang, Chen, Zhang, Peng, Li, Xie, Nong, et~al.]{jiang2024megascale}
Ziheng Jiang, Haibin Lin, Yinmin Zhong, Qi~Huang, Yangrui Chen, Zhi Zhang, Yanghua Peng, Xiang Li, Cong Xie, Shibiao Nong, et~al.
\newblock $\{$MegaScale$\}$: Scaling large language model training to more than 10,000 $\{$GPUs$\}$.
\newblock In \emph{21st USENIX Symposium on Networked Systems Design and Implementation (NSDI 24)}, pages 745--760, 2024.

\bibitem[Korthikanti et~al.(2023)Korthikanti, Casper, Lym, McAfee, Andersch, Shoeybi, and Catanzaro]{korthikanti2023reducing}
Vijay~Anand Korthikanti, Jared Casper, Sangkug Lym, Lawrence McAfee, Michael Andersch, Mohammad Shoeybi, and Bryan Catanzaro.
\newblock Reducing activation recomputation in large transformer models.
\newblock \emph{Proceedings of Machine Learning and Systems}, 5, 2023.

\bibitem[Lepikhin et~al.(2020)Lepikhin, Lee, Xu, Chen, Firat, Huang, Krikun, Shazeer, and Chen]{lepikhin2020gshard}
Dmitry Lepikhin, HyoukJoong Lee, Yuanzhong Xu, Dehao Chen, Orhan Firat, Yanping Huang, Maxim Krikun, Noam Shazeer, and Zhifeng Chen.
\newblock Gshard: Scaling giant models with conditional computation and automatic sharding.
\newblock \emph{arXiv preprint arXiv:2006.16668}, 2020.

\bibitem[Li et~al.(2024)Li, Fu, Ge, Lin, Wang, Niu, Wang, Zhang, Nie, and Cui]{li2024malleus}
Haoyang Li, Fangcheng Fu, Hao Ge, Sheng Lin, Xuanyu Wang, Jiawen Niu, Yujie Wang, Hailin Zhang, Xiaonan Nie, and Bin Cui.
\newblock Malleus: Straggler-resilient hybrid parallel training of large-scale models via malleable data and model parallelization.
\newblock \emph{arXiv preprint arXiv:2410.13333}, 2024.

\bibitem[Liu et~al.(2023)Liu, Zaharia, and Abbeel]{liu2023ringattentionblockwisetransformers}
Hao Liu, Matei Zaharia, and Pieter Abbeel.
\newblock Ring attention with blockwise transformers for near-infinite context, 2023.
\newblock URL \url{https://arxiv.org/abs/2310.01889}.

\bibitem[Luo et~al.(2019)Luo, Lin, Zhuo, and Qian]{luo2019hop}
Qinyi Luo, Jinkun Lin, Youwei Zhuo, and Xuehai Qian.
\newblock Hop: Heterogeneity-aware decentralized training.
\newblock In \emph{Proceedings of the Twenty-Fourth International Conference on Architectural Support for Programming Languages and Operating Systems}, pages 893--907, 2019.

\bibitem[mega()]{megatron}
mega.
\newblock Megatron-lm.
\newblock \url{https://github.com/NVIDIA/Megatron-LM}, 2020.

\bibitem[Narayanan et~al.(2021)Narayanan, Shoeybi, Casper, LeGresley, Patwary, Korthikanti, Vainbrand, Kashinkunti, Bernauer, Catanzaro, et~al.]{narayanan2021efficient}
Deepak Narayanan, Mohammad Shoeybi, Jared Casper, Patrick LeGresley, Mostofa Patwary, Vijay Korthikanti, Dmitri Vainbrand, Prethvi Kashinkunti, Julie Bernauer, Bryan Catanzaro, et~al.
\newblock Efficient large-scale language model training on gpu clusters using megatron-lm.
\newblock In \emph{Proceedings of the International Conference for High Performance Computing, Networking, Storage and Analysis}, pages 1--15, 2021.

\bibitem[Ng et~al.(2023)Ng, Demoulin, and Liu]{ng2023paella}
Kelvin~KW Ng, Henri~Maxime Demoulin, and Vincent Liu.
\newblock Paella: Low-latency model serving with software-defined gpu scheduling.
\newblock In \emph{Proceedings of the 29th Symposium on Operating Systems Principles}, pages 595--610, 2023.

\bibitem[Ousterhout et~al.(2015)Ousterhout, Rasti, Ratnasamy, Shenker, and Chun]{ousterhout2015making}
Kay Ousterhout, Ryan Rasti, Sylvia Ratnasamy, Scott Shenker, and Byung-Gon Chun.
\newblock Making sense of performance in data analytics frameworks.
\newblock In \emph{12th USENIX Symposium on Networked Systems Design and Implementation (NSDI 15)}, pages 293--307, 2015.

\bibitem[Rajbhandari et~al.(2020)Rajbhandari, Rasley, Ruwase, and He]{rajbhandari2020zero}
Samyam Rajbhandari, Jeff Rasley, Olatunji Ruwase, and Yuxiong He.
\newblock Zero: Memory optimizations toward training trillion parameter models.
\newblock In \emph{SC20: International Conference for High Performance Computing, Networking, Storage and Analysis}, pages 1--16. IEEE, 2020.

\bibitem[Ramamoorthy et~al.(2020)Ramamoorthy, Das, and Tang]{ramamoorthy2020straggler}
Aditya Ramamoorthy, Anindya~Bijoy Das, and Li~Tang.
\newblock Straggler-resistant distributed matrix computation via coding theory: Removing a bottleneck in large-scale data processing.
\newblock \emph{IEEE Signal Processing Magazine}, 37\penalty0 (3):\penalty0 136--145, 2020.

\bibitem[Tao et~al.(2024)Tao, Liu, Dou, Muennighoff, Wan, Luo, Lin, and Wong]{tao2024scaling}
Chaofan Tao, Qian Liu, Longxu Dou, Niklas Muennighoff, Zhongwei Wan, Ping Luo, Min Lin, and Ngai Wong.
\newblock Scaling laws with vocabulary: Larger models deserve larger vocabularies.
\newblock \emph{arXiv preprint arXiv:2407.13623}, 2024.

\bibitem[Team({\natexlab{a}})]{contextparallel}
NVIDIA Team.
\newblock Context paralellism.
\newblock \url{"https://docs.nvidia.com/megatron-core/developer-guide/latest/api-guide/context_parallel.html"}, {\natexlab{a}}.

\bibitem[Team({\natexlab{b}})]{cpythongc}
Python Team.
\newblock Garbage collector design.
\newblock \url{"https://github.com/python/cpython/blob/main/InternalDocs/garbage_collector.md"}, {\natexlab{b}}.

\bibitem[Team({\natexlab{c}})]{torchtcpstore}
PyTorch Team.
\newblock Pytorch distributed key-value store -- tcp store.
\newblock \url{"https://pytorch.org/docs/stable/distributed.html#torch.distributed.TCPStore"}, {\natexlab{c}}.

\bibitem[Team(2024)]{imbue}
The~Imbue Team.
\newblock From bare metal to a 70b model: infrastructure set-up and scripts.
\newblock \url{"https://imbue.com/research/70b-infrastructure/"}, 2024.

\bibitem[Wei et~al.(2024)Wei, Zhu, Zhao, Cheng, Li, L{\"u}, Cheng, Zhang, Zhang, Zeng, et~al.]{wei2024skywork}
Tianwen Wei, Bo~Zhu, Liang Zhao, Cheng Cheng, Biye Li, Weiwei L{\"u}, Peng Cheng, Jianhao Zhang, Xiaoyu Zhang, Liang Zeng, et~al.
\newblock Skywork-moe: A deep dive into training techniques for mixture-of-experts language models.
\newblock \emph{arXiv preprint arXiv:2406.06563}, 2024.

\bibitem[{Wikipedia contributors}()]{wikimpp}
{Wikipedia contributors}.
\newblock Multiway number partitioning --- {W}ikipedia.
\newblock \url{"https://en.wikipedia.org/wiki/Multiway_number_partitioning"}.

\bibitem[Wu et~al.(2024)Wu, Wang, Yu, Yang, Wu, Duan, Yang, Wang, Qu, and Zhang]{wu2024falcon}
Tianyuan Wu, Wei Wang, Yinghao Yu, Siran Yang, Wenchao Wu, Qinkai Duan, Guodong Yang, Jiamang Wang, Lin Qu, and Liping Zhang.
\newblock {FALCON}: Pinpointing and mitigating stragglers for large-scale hybrid-parallel training.
\newblock \emph{arXiv preprint arXiv:2410.12588}, 2024.

\bibitem[Yeung et~al.(2024)Yeung, Qi, Lin, and Wan]{yeung2024balancing}
Man~Tsung Yeung, Penghui Qi, Min Lin, and Xinyi Wan.
\newblock Balancing pipeline parallelism with vocabulary parallelism.
\newblock \emph{arXiv preprint arXiv:2411.05288}, 2024.

\bibitem[Zhang et~al.(2024)Zhang, Zhong, Ming, Hu, Sun, Ge, Zhu, and Jin]{zhang2024disttrainaddressingmodeldata}
Zili Zhang, Yinmin Zhong, Ranchen Ming, Hanpeng Hu, Jianjian Sun, Zheng Ge, Yibo Zhu, and Xin Jin.
\newblock Dist{T}rain: Addressing model and data heterogeneity with disaggregated training for multimodal large language models, 2024.
\newblock URL \url{https://arxiv.org/abs/2408.04275}.

\bibitem[Zhao et~al.(2023)Zhao, Gu, Varma, Luo, Huang, Xu, Wright, Shojanazeri, Ott, Shleifer, Desmaison, Balioglu, Damania, Nguyen, Chauhan, Hao, Mathews, and Li]{zhao2023pytorchfsdp}
Yanli Zhao, Andrew Gu, Rohan Varma, Liang Luo, Chien-Chin Huang, Min Xu, Less Wright, Hamid Shojanazeri, Myle Ott, Sam Shleifer, Alban Desmaison, Can Balioglu, Pritam Damania, Bernard Nguyen, Geeta Chauhan, Yuchen Hao, Ajit Mathews, and Shen Li.
\newblock Pytorch fsdp: Experiences on scaling fully sharded data parallel.
\newblock \emph{Proc. VLDB Endow.}, 16\penalty0 (12):\penalty0 3848–3860, August 2023.
\newblock ISSN 2150-8097.
\newblock \doi{10.14778/3611540.3611569}.
\newblock URL \url{https://doi.org/10.14778/3611540.3611569}.

\end{thebibliography}

\end{document}